\documentclass[%
 reprint,
 superscriptaddress,
 showpacs,
 amsmath,amssymb,
 aps,
 pra,
 longbibliography,
]{revtex4-1}

\usepackage{graphicx}
\usepackage{braket}
\usepackage{bm}
\usepackage{times}
\usepackage{txfonts}
\usepackage{amsmath}
\usepackage{mathrsfs}
\allowdisplaybreaks[4]
\usepackage{xcolor}
\usepackage{color}
\usepackage[colorlinks=true,
linkcolor=blue,
anchorcolor=blue,
citecolor=blue,
urlcolor=blue
]{hyperref}

\begin{document}


\title{Magnetic-Field effect in Laser-assisted XUV ionization}

\author{Jintai Liang}
\affiliation{School of Physics and Wuhan National Laboratory for Optoelectronics, Huazhong University of Science and Technology, Wuhan 430074, China }
\author{Yueming Zhou}
\email{zhouymhust@hust.edu.cn}
\affiliation{School of Physics and Wuhan National Laboratory for Optoelectronics, Huazhong University of Science and Technology, Wuhan 430074, China }
\author{Wei-Chao Jiang}
\email{jiang.wei.chao@szu.edu.cn}
\affiliation{College of Physics And Optoelectronic Engineering, Shenzhen University, Shenzhen 518060, China}
\author{Min Li}
\affiliation{School of Physics and Wuhan National Laboratory for Optoelectronics, Huazhong University of Science and Technology, Wuhan 430074, China }
\author{Peixiang Lu}
\affiliation{School of Physics and Wuhan National Laboratory for Optoelectronics, Huazhong University of Science and Technology, Wuhan 430074, China }
\affiliation{Optics Valley Laboratory, Hubei 430074, China}

\begin{abstract}
  The magnetic-field effect of the laser pulse is investigated in laser-assisted XUV ionization. By numerically solving the three-dimensional time-dependent Schr\"odinger equation, we find that the photoelectron momentum distribution is distorted by the magnetic-field effect of the IR streaking field. It results in a transverse-momentum- and time-delay-dependent longitudinal photoelectron momentum shift, which is further analytically confirmed by employing the nondipole corrected strong field approximation. We also demonstrate that the magnetic field does not affect the time shift retrieved from the streaking spectrum, although the streaking spectrum is apparently altered by the momentum shift induced by the magnetic-field effect. Our work reveals the time-resolved nondipole effect of the laser pulse by the attosecond streaking technique.
\end{abstract}


\maketitle

\section{introduction}
The electric dipole approximation is widely used to facilitate calculations and understanding of strong-field ionization, in which the magnetic component of the laser field is neglected. It usually holds well for the most commonly used laser sources and intensities. This is because the ratio of the amplitudes for electric field to magnetic field equals the speed of light $c$, and thus the photoelectron dynamics are mainly determined by the electric field.  While in the high-intensity long-wavelength limit \cite{Milo_evi__2006,Reiss2008}, the effects of the magnetic fields are non-negligible. 

With recent advances in detecting technologies, the magnetic-field effect of the laser pulse has become observable and aroused considerable interest. Specifically, the magnetic effect shows up as an asymmetric photoelectron momentum distribution (PEMD) along the laser propagation direction ($y$ axis in our work). The asymmetric PEMD results in a nonzero expectation value of $p_y$ representing the linear momentum transfer from photons to photoelectrons. In the past decade, the manifestation of this nonzero $\braket{p_y}$ in various ionization regimes has been deeply studied in experiment \cite{Smeenk2011,Ludwig2014,Hartung2019,Lin2022,Lin2022_1,Sun2020,Chen2020,Grundmann2020} and theory \cite{Klaiber2013,Titi2012,Jensen2020,Chelkowski2014,Chelkowski2017,Chelkowski2015,He2017,Liu2013,Wang2017} (see Ref.\;\cite{Maurer2021} for recent review). A more interesting issue is the time-resolved momentum transfer from photon to photoelectrons. It provides the information of the instantaneous nondipole effect on the photoelectrons. To reveal this effect, a measurement with the subcycle resolution is needed. Very recently, the subcycle time-resolved momentum transfer measurement in tunneling ionization has been achieved by an attoclock protocol \cite{Willenberg2019,Hongcheng2020}. 

Similar to the attoclock, the attosecond streaking technique, which is based on an attosecond extreme ultraviolet (XUV) pulse serving as the pump pulse and a phase controlled infrared (IR) field as the probe pulse, is also capable of delivering real-time information on electronic processes in ultrafast time scales.  In attosecond streaking scheme, the electrons are released by the XUV pulse and accelerated by the IR laser field. The PEMDs are shifted by the amount of momentum transferred from the IR field to the continuum electron wave packets (EWPs). Thus the time information is mapped onto the energy axis of photoelectrons with attosecond precision. Due to its subcycle resolution, this technique has shown extensive applications, such as characterizing the attosecond pulse as well as the IR field \cite{Goulielmakis2004,Itatani2002,Kienberger2004,Kitzler2002}, and observing the buildup of the Fano resonance in the time domain \cite{Wang2010,Wickenhauser2005,Ning2014}. Therefore, the attosecond streaking technique has great potential to investigate the time-resolved nondipole effect of the laser pulse \cite{Jensen2020,Maurer2021}. Yet, to the best of our knowledge, few works have studied the nondipole effect in the attosecond streaking. 

Moreover, since the time delay in photoemission from the $2s$ and the $2p$ states was measured in the experiment \cite{Schultze2010}, the time shift retrieved by the attosecond streaking technique has triggered broad interesting in atomic \cite{Kheifets2010,Baggesen2010,Ivanov2011,Pazourek2012,Su2014,Saalmann2020,Pazourek2013} and molecular ionization \cite{Ivanov2012,Gopal2013,Serov2013}, and the photoemission from the solid surface \cite{Ossiander2018,Zhang2009,Liao2014} (see Ref.\;\cite{Pazourek2015} for review).  The time shift can be roughly divided into two parts. One is the intrinsic time delay that is independent on the probe pulse, such as the Wigner time delay in atomic photoionization \cite{Pazourek2015}. The other part is the time shift induced by the IR-probe pulse. For instance, the Coulomb-laser coupling \cite{Ivanov2011,Pazourek2013,Pazourek2015} (CLC) and the dipole-laser coupling \cite{Baggesen2010} (dLC) time shift. These time shifts are obtained by applying the dipole approximation. Previous work has demonstrated that the nondipole effect induced by the XUV pulse does not affect the intrinsic atomic time delay \cite{Spiewanowski2012}. However, the magnetic-field effect of the IR streaking field on the time shift is still an open question. 

In this work, we focus on revealing the magnetic-field effect of the streaking field in laser-assisted XUV photoionization. By numerically solving the three-dimensional (3D) time-dependent Schr\"odinger equation (TDSE), the time-resolved linear momentum transfer from photon to the photoelectrons in the laser-assisted XUV photoionization is obtained. It shows that the time-delay-dependent linear momentum transfer has two contributions, the nondipole effect of the XUV pulse and the magnetic-field effect of the IR field. To reveal the magnetic-field effect of the IR pulse in laser-assisted XUV ionization, the manifestation of the nondipole effect on the PEMDs is investigated. We find that the PEMDs are distorted by the magnetic-field effect of the streaking field, which results in a transverse-momentum- and time-delay-dependent longitudinal momentum shift. This momentum shift is further analytically confirmed by employing the nondipole corrected strong-field approximation (ndSFA). We also demonstrate that the magnetic field does not affect the time shift retrieved by the attosecond streaking technique, although the momentum shift induced by the magnetic-field effect apparently alters the photoelectron streaking spectrum. 

\section{numerically solving 3D-TDSE}
The dynamics of an atom interacting with the laser pulses are governed by the TDSE, in which the Hamiltonian beyond the dipole approximation by including corrections to the first order in $1/c$ is given by (atomic units are used unless otherwise stated) \cite{Hartung2019,Brennecke_2018}
\begin{equation}\label{Eq:Hamitoni}
  \begin{aligned}
    H=&\frac{1}{2}\left[{\bf{p}}+{\bf{A}}(t)+\frac{{\bf{e}}_y}{c}\left({\bf{p}}\cdot{\bf{A}}(t)+\frac{1}{2}{\bf{A}}^2(t)\right)\right]^2\\
    +&V\left({\bf{r}}-\frac{y}{c}{\bf{A}}(t)\right),
  \end{aligned}
\end{equation}
where ${\bf{A}}(t)={\bf{A}}(t,y=0)$ is the laser vector potential at the position of the nucleus, and $V({\bf{r}})=-1/r$ is the Coulomb potential of the H atom. To speed up the time propagation, we expand the shifted potential to first order in $1/c$, i.e. \cite{Brennecke_2018},
\begin{equation}
  V\left({\bf{r}}-\frac{y}{c}{\bf{A}}(t)\right)\approx V({\bf{r}})-\frac{y}{c}{\bf{A}}(t)\cdot\nabla V({\bf{r}}).
\end{equation}
Then for a linearly polarized laser pulse along the $z$ axis, the Hamiltonian in Eq.\;(\ref{Eq:Hamitoni}) is reduced to
\begin{equation}\label{Eq:Hamiton_reduce}
  \begin{aligned}
    H=&\frac{1}{2}{\bf{p}}^2+p_zA(t)+\frac{1}{c}p_y\left(p_zA(t)+\frac{1}{2}A^2(t)\right)\\
      -&\frac{1}{r}-\frac{yz}{cr^3}{{A}}(t).
  \end{aligned}
\end{equation}
Note that here the purely  time-dependent quadratic $\frac{1}{2}A^2(t)$ term has been removed by the gauge transformation
\begin{equation}
  \Psi'=\exp\left[i\int_0^t\frac{1}{2}A^2(t')dt'\right]\Psi.
\end{equation}

In this work, the laser vector potential is written as  
\begin{equation}
  A(t)=A_{\rm XUV}(t-\tau)+A_{\rm IR}(t),
\end{equation}
where,
\begin{equation}
    A_{\rm XUV}(t)=A^0_{\rm XUV}\exp\left[-2\ln2(\frac{t}{5T_{\rm XUV}})^2\right]\cos(\omega_{\rm XUV}t),
\end{equation}
and 
\begin{equation}
    A_{\rm IR}(t)=A^0_{\rm IR}\cos\left(\pi\frac{t}{2T_{\rm IR}}\right)^2\sin(\omega_{\rm IR}t),
\end{equation}
are the vector potentials of the XUV pulse and IR pulse, respectively. The XUV and IR pulses are both linearly polarized along the $z$ axis and propagate along the $y$ axis. $\tau$ is the time delay between the IR field and the XUV pulse. $A^0_{\rm XUV,IR}$, $T_{\rm IR,XUV}$ and $\omega_{\rm XUV,IR}$ are the amplitude of the vector potential, period and center frequency of laser pulse. In our calculation, unless otherwise stated, the frequency of the XUV pulse is $\omega_{\rm XUV}=2.5$ a.u. and its intensity is 1$\times$10$^{13}$ W/cm$^2$. The wavelength and intensity of the IR field are 3600 nm and 5$\times$10$^{12}$ W/cm$^2$, respectively.

In our simulation, the TDSE with the Hamiltonian in Eq.\;(\ref{Eq:Hamiton_reduce}) is solved in the spherical coordinates, in which the wavefunction $\Psi({\bf{r}},t)$ is expanded by spherical harmonics $\ket{l,m}$,
\begin{equation}
  \ket{\Psi({\bf{r}},t)}=\sum_{l,m}\frac{R_{l,m}(r,t)}{r}\ket{l,m},
\end{equation}
where $R_{l,m}(r,t)$ is the radial part of the wavefunction. This radial wavefunction is discretized by the finite-element discrete variable representation (FE-DVR) method \cite{Liang2022,Liang:21,Liang_2020}. The angular quantum number $l$ and magnetic quantum number $m$ are chosen up to 100 and 10, respectively. The time propagation of the TDSE is calculated by the split-Lanczos method with the time step fixed at $\Delta t=0.01$ a.u.. The maximal box size for the radial coordinate is chosen to be 200 a.u.. An absorbing function has been applied in each step of time propagation of the wavefunction, which is written as $F(r)=1-1/(1+e^{(r-R_c)/L})$ with $R_c=150$ a.u. and $L=2$ a.u.. The wavefunction $\Psi({\bf{r}},t)$ is split into the inner part $\Psi_{\rm in}({\bf{r}},t)=\Psi({\bf{r}},t)F(r)$ and the outer part $\Psi_{\rm out}=\Psi({\bf{r}},t)-\Psi_{\rm in}({\bf{r}},t)$ by the absorbing function. The inner wavefunction evolves strictly as TDSE, while the outer part $\Psi_{\rm out}$ is propagated by Coulomb-Volkov propagator \cite{Arobo2008}. {At each time step $t_i$, $\Psi_{\rm out}({\bf{r}},t_i)$ is projected to the scattering state $\Psi_{{p}}({\bf r})$ to obtain the ionization amplitude $\mathcal{M}'({\bf{p}},t_i)=\braket{\Psi_{{p}}({\bf r})|\Psi_{\rm out}({\bf{r}},t_i)}$. Then from time $t_i$ to the next time step $t_{i+1}=t_{i}+\Delta t$, the ionization continuum is only changed by a Volkov phase with the nondipole correction \cite{Simon2021}
\begin{equation}
  U_{{\bf{p}}}(t_i,t_{i+1})=e^{-i\int_{t_i}^{t_{i+1}}[\frac{p^2}{2}+{\bf{p}}\cdot{\bf{A}}(\tau)+\frac{p_y}{c}({\bf{p}}\cdot{\bf{A}}(\tau)+{\bf{A}}^2(\tau)/2)]d\tau}. 
\end{equation}
At time $t_{i+1}$, we add the amplitude propagated from $t_i$ and the new splitting amplitude $\mathcal{M}'({{\bf{p}},t_{i+1}})$,
\begin{equation}
  \mathcal{M}({\bf{p}},t_{i+1})=U_{{\bf{p}}}(t_i,t_{i+1})\mathcal{M}({\bf{p}},t_i)+\mathcal{M}'({\bf{p}},t_{i+1}). 
\end{equation}
Note that we use $\mathcal{M}'$ to indicate the new splitting amplitude from the TDSE calculation and $\mathcal{M}$ to indicate the sum of all amplitudes including the new splitting one and those propagated from the previous splitting times. 
At the end time $t_f$ of the propagation, splitting is not needed any more and the new produced amplitude $\mathcal{M}'({\bf{p}},t_f)$ is extracted from the whole wave function $\Psi({\bf{r}},t_f)$ at time $t_f$. Adding all the amplitudes at time $t_f$, we obtain the final ionization amplitude $\mathcal{M}({\bf{p}},t_f)$. In our work, $\Psi_{{p}}({\bf r})$ is chosen as the scattering state of H atom \cite{Jiang2017}, which is normalized by $\int d{\bf{r}}\Psi^*_{\bm{p}}({\bf{r}})\Psi_{\bm{p'}}({\bf{r}})=\delta({\bm{p}}-{\bm{p'}})$.
The convergence of our calculations has been confirmed by changing $l$, $m$ and $R_c$ in our calculations. The initial wavefunction is prepared by the imaginary-time propagation which is chosen as the ground state of H atom.} Then the photoelectron average momentum $\braket{p_y}$ is obtained by
\begin{equation}
  \braket{p_y}=\frac{\iiint p_r\sin\theta\sin\phi|\mathcal{M}({\bm{p}})|^2 p_r^2\sin\theta dp_r d\theta d\phi }{\iiint |\mathcal{M}({\bm{p}})|^2 p_r^2\sin\theta dp_r d\theta d\phi}. 
\end{equation}

\section{Results and Discussion}

\begin{figure}[t]
  \includegraphics[width=0.45\textwidth]{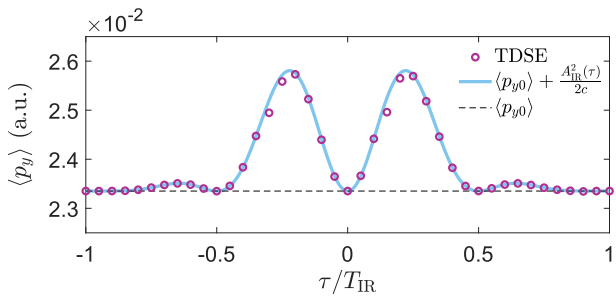}
  \caption{Average photoelectron momentum $\braket{p_y}$ as a function of the time delay between the XUV and IR pulse. The circles and the solid line represent the results obtained by numerically solving the TDSE and predicted by Eq.\;(\ref{Eq:classical_py_ave}). The dashed line shows the linear momentum transfer by the XUV pulse, i.e., $\braket{p_{y0}}=8(\omega_{\rm XUV}-I_p)/(5c)$.}
  \label{fig:figure1}
\end{figure}

\subsection{Linear momentum transfer in laser-assisted XUV photoionization}
By numerically solving the TDSE, the linear momentum transfer $\braket{p_y}$ in the laser-assisted XUV photoionization is obtained, as displayed by the circles in Fig.\;\ref{fig:figure1}. It shows that the average momentum oscillates with the time delay between the XUV and the IR pulse, which can be understood by the classical analysis of the photoelectrons dynamics. Mechanistically, after ionization, the photoelectrons are accelerated by the Lorentz force ${\bf{F}}_L=-({\bf{E}}+{\bf{V}}\times{\bf{B}})$, where ${\bf{E}}$ and ${\bf{B}}$ are the electric and magnetic field of the laser pulse. According to Newton’s equation neglecting the Coulomb potential, the final momentum along the propagation direction of the photoelectrons is given by \cite{Willenberg2019}
\begin{equation}\label{Eq:classical}
  p_{yf}=p_{y0}+\frac{1}{2c}{\bf{A}}(t_0)\cdot\left[{\bf{A}}(t_0)-2{\bf{p}}_{\perp0}\right],
\end{equation}
where $p_{y0}$ and ${\bf{p}}_{\perp0}$ denote the electron's initial momentum along the propagation direction and in the polarization plane, respectively. ${\bf{A}}(t_0)$ is the vector potential at the ionization time $t_0$. It implies that the linear momentum transfer in the laser-assisted XUV ionization can be expressed as 
\begin{equation}\label{Eq:classical_py_ave}
  \braket{p_y}=\braket{p_{y0}}+\frac{A^2_{\rm IR}(\tau)}{2c},
\end{equation}
where the relations $\braket{{\bf{p}}_{\perp0}}=0$ and $\tau=t_0$ are used. The initial average momentum $\braket{p_{y0}}$ represents the linear momentum transfer from the XUV pulse given by $\braket{p_{y0}}=8(\omega_{\rm XUV}-I_p)/(5c)$ for the $1s$ state of the H atom \cite{Chelkowski2014}. The solid line in Fig.\;\ref{fig:figure1} represents the results obtained by Eq.\;(\ref{Eq:classical_py_ave}), which agrees well with the results obtained by numerically solving the TDSE. Eq.\;(\ref{Eq:classical_py_ave}) indicates that the linear momentum transfer in the laser-assisted XUV photoionization has two contributions. The time-delay-independent part originates from nondipole effect of the XUV pulse, and the time-delay-dependent component is attributed to the magnetic-field effect of the IR pulse on the ionized photoelectrons. 

\begin{figure*}[t]
  \includegraphics[width=0.9\textwidth]{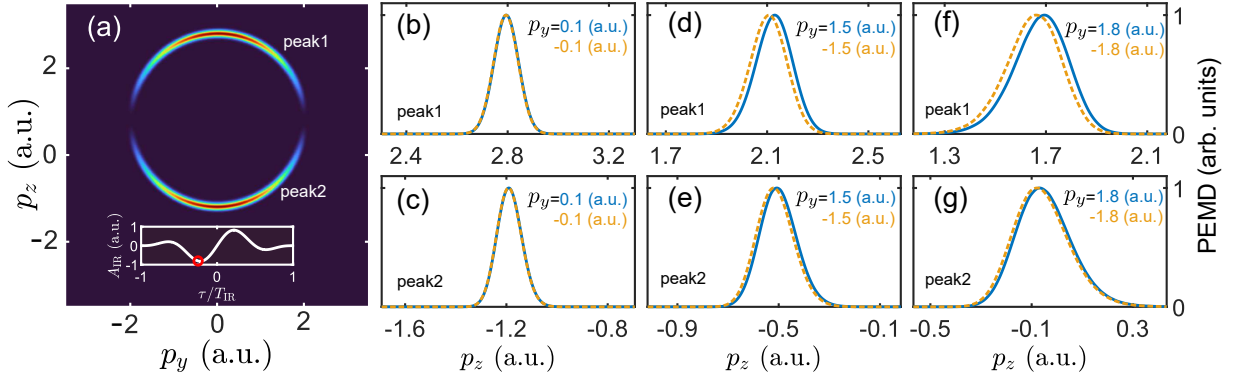}
  \caption{(a) PEMD in ($p_y$, $p_z$) plane from single-photon ionization in the XUV-IR pulse obtained by numerically solving the TDSE. The inset shows the vector potential of the IR pulse and the red circle marks the time delay between XUV and IR pulse. (b)-(g) Cuts of the PEMD at $|p_y|$=0.1 (b,c), $|p_y|$=1.5 (d,e) and $|p_y|$=1.8 (f,g). Note that the cuts are normalized by their maxima. The top and bottom rows display the PEMDs corresponding to the peak1 and peak2, respectively.}
  \label{fig:figure2}
\end{figure*}

\subsection{Magnetic-field effect of the IR pulse on the PEMDs}
\label{sec:mag_eff_IR}

To reveal the magnetic-field effect of the IR pulse in the laser-assisted XUV ionization, the manifestation of the nondipole effect on the PEMDs is investigated. Figure\;\ref{fig:figure2}(a) displays the obtained PEMD in the ($p_y$, $p_z$) plane (i.e., $p_x$=0) with $\tau$=$-T_{\rm IR}/4$. The PEMD shows a ring-like structure with two half rings locating at $p_z>0$ and $p_z<0$ plane denoted by peak1 and peak2, respectively. The cuts of the PEMDs at various $|p_y|$ are displayed in Figs.\;\ref{fig:figure2}(b)-\ref{fig:figure2}(g). Note that the cuts are normalized by their maxima. It shows that there is a peak shift between the PEMD for $p_y>0$ and $p_y<0$. This peak shift increases with $p_y$ as indicated by Figs.\;\ref{fig:figure2}(b), \ref{fig:figure2}(d) and \ref{fig:figure2}(f). More interestingly, the peak shift of the peak1 is much larger than that for peak2, as shown by Figs.\;\ref{fig:figure2}(f) and \ref{fig:figure2}(g). 

\begin{figure}[b]
  \includegraphics[width=0.45\textwidth]{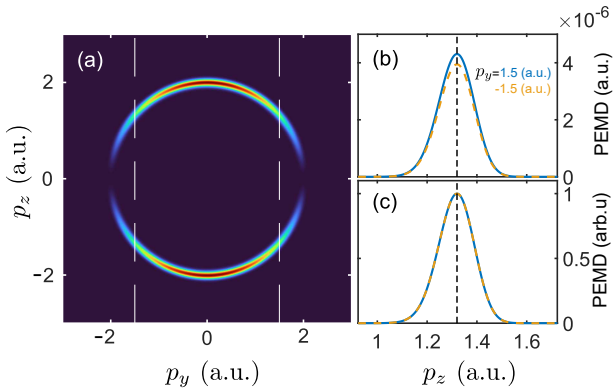}
  \caption{(a) PEMD in ($p_y$, $p_z$) plane from ionization in XUV pulse alone obtained by numerically solving TDSE. (b) Cuts of the PEMD at $p_y$=1.5 (blue solid line) and $p_y$=-1.5 (yellow dashed line). (c) Normalized PEMD by their maximum. The colors of lines correspond to the colors in (b).}
  \label{fig:figure3}
\end{figure}

The PEMD from ionization in the XUV pulse alone is also obtained by numerically solving TDSE, as displayed in Fig.\;\ref{fig:figure3}(a). The cuts of the PEMD at $p_y=\pm 1.5$ a.u. are displayed in Fig.\;\ref{fig:figure3}(b). It shows that the yield of the PEMDs are different for $p_y=\pm 1.5$ a.u., which is responsible for the linear momentum transfer by XUV pulse, i.e., $\braket{p_{y0}}$ in Eq.\;(\ref{Eq:classical_py_ave}). The normalized PEMDs are shown in Fig.\;\ref{fig:figure3}(c). No momentum shift is observed for the ionization by the XUV pulse alone. It indicates that the momentum shift shown in Fig.\;\ref{fig:figure2} originates from the magnetic-field effect of the IR pulse. 

To analytically explore the origination of the peak shift, we extend the strong-field approximation (SFA) to include nondipole effects (ndSFA) in attosecond streaking, as addressed in Appendix\;\ref{Sec:appB}. In ndSFA, the transition amplitude for the ionization by a XUV pulse in the presence of IR streaking field is written as
\begin{equation}{\label{Eq:ndSFA:MP_d}}
  M^{\rm SFA}_{\rm nd}({\bf{p}},\tau)\approx-i\int_{0}^{\infty}dt\langle {\bf{k}}(t)|{\bf E}_{\rm XUV}(t-\tau)\cdot{\bf{r}}|\psi_0\rangle e^{-iS_{\rm nd}({\bf{p}},t)}.
\end{equation}
Here $S_{\rm nd}({\bf{p}},t)=-I_pt+\int_t^{\infty}d\xi{{\bf{k}}^2(\xi)}/2$, and ${\bf{k}}(t)={\bf{p+A_{\rm IR}}}(t)+\frac{{\bf{e}}_y}{c}({\bf{p}}\cdot{\bf{A}}_{\rm IR}(t)+\frac{1}{2}{\bf{A}}_{\rm IR}^2(t))$. ${\bf E}_{\rm XUV}(t)=-d{\bf A}_{\rm XUV}(t)/dt$ is the electric field of the XUV pulse. With the saddle point approximation (shown in Appendix\;\ref{Sec:appB}), the nondipole modified attosecond streaking camera with $p_x=0$ is expressed as
\begin{equation}\label{Eq:atto-carma_nondipole_y_d}
  [p_z+A_{\rm IR}(\tau)]^2+\left[p_y+\frac{p_z}{c}A_{\rm IR}(\tau)+\frac{A_{\rm IR}^2(\tau)}{2c}\right]^2=k_0^2.
\end{equation}
Here $k_0=\sqrt{2(\omega_{\rm XUV}-I_p)}$. While within the dipole approximation, the streaking camera is given by 
\begin{equation}\label{Eq:atto-carma_dipole_y_d}
    [p_z^d+A_{\rm IR}(\tau)]^2+p_y^2=k_0^2.
\end{equation}
Eq.\;(\ref{Eq:atto-carma_nondipole_y_d}) indicates that the PEMDs are distorted by the magnetic-field effect of the streaking field. 

\begin{figure}[t]
  \includegraphics[width=0.3\textwidth]{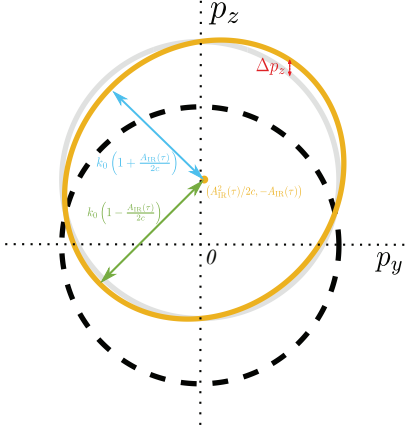}
  \caption{Schematic illustration of the effect of magnetic field of the IR field on the PEMDs. The black solid line represents the PEMD from single-photon ionization in the XUV pulse alone. It is a ring centered at $(p_y,p_z)=(0,0)$ with the radius of $k_0$. The yellow solid line represent the PEMD in the XUV-IR pulse with nondipole correction determined by Eq.\;(\ref{Eq:atto-carma_nondipole_y_d}). It is a ellipse centered at $(p_y,p_z)=(A^2_{\rm IR}(\tau)/2c,-A_{\rm IR}(\tau))$ with major and minor axes $k_0[1-A_{\rm IR}(\tau)/2c]$ and $k_0[1+A_{\rm IR}(\tau)/2c]$, respectively. The PEMD with the dipole approximation is also displayed by the gray solid line for comparison, which is a ring centered at $\{0,-A_{\rm IR}(\tau)\}$.}
  \label{fig:figure4}
\end{figure}

Figure\;\ref{fig:figure4} schematically illustrates the distortion of the PEMDs by the magnetic-field effect. The PEMD for the ionization of the XUV pulse alone is a ring-like structure centered at $(p_y,p_z)=(0,0)$. Within the effect of the electric field of the IR pulse, the PEMD shift along the $p_z$ axis with the momentum of $-A_{\rm IR}(\tau)$. When the magnetic effect is considered, the PEMD is distorted showing as an ellipse-like structure, as indicated by Eq.\;(\ref{Eq:atto-carma_nondipole_y_d}). The ellipse is centered at $(p_y,p_z)=(A^2_{\rm IR}(\tau)/2c,-A_{\rm IR}(\tau))$ with major and minor axes $k_0[1-A_{\rm IR}(\tau)/2c]$ and $k_0[1+A_{\rm IR}(\tau)/2c]$, respectively. Therefore, for the same transverse momentum $p_y$, there is a longitudinal momentum shift compared with the PEMD within the dipole approximation, which is denoted by $\Delta p_z$ as shown in Fig.\;\ref{fig:figure4}. This longitudinal momentum shift induced by the magnetic field is responsible for the phenomena described in Fig.\;\ref{fig:figure2}. 

According to the above analysis, the longitudinal photoelectron momentum can be written as $p_z=p_z^d+\Delta p_z$, where $p_z^d$ is the longitudinal momentum with the dipole approximation shown in Eq.\;(\ref{Eq:atto-carma_dipole_y_d}). Then Eq.\;(\ref{Eq:atto-carma_nondipole_y_d}) is written as 
\begin{equation}
  [p_z^d+\Delta p_z+A_{\rm IR}(\tau)]^2+\left[p_y+\frac{p_z^d+\Delta p_z}{c}A_{\rm IR}(\tau)+\frac{A_{\rm IR}^2(\tau)}{2c}\right]^2=k_0^2.
\end{equation}
Within the order of $1/c$, this equation is reduced to 
\begin{equation}\label{Eq:ana_porcudure}
  \begin{aligned}
    &[p_z^d+A_{\rm IR}(\tau)]^2+2\Delta p_y[p_z^d+A_{\rm IR}(\tau)]+p_y^2\\
    +&\frac{2p_yp_z^d}{c}A_{\rm IR}(\tau)+\frac{p_yA_{\rm IR}^2(\tau)}{c}=k_0^2.
  \end{aligned}
\end{equation}
Inserting Eq.\;(\ref{Eq:atto-carma_dipole_y_d}) into Eq.\;(\ref{Eq:ana_porcudure}), we obtain 
\begin{equation}
  \frac{2p_yp_z^d}{c}A_{\rm IR}(\tau)+\frac{p_yA_{\rm IR}^2(\tau)}{c}+2\Delta p_z[p_z^d+A_{\rm IR}(\tau)]=0.
\end{equation}
Then the longitudinal momentum shift $\Delta p_z$ is expressed as
\begin{equation}
  \Delta p_z=-p_y\frac{[2p_z^d+A_{\rm IR}(\tau)]A_{\rm IR}(\tau)}{2c[p_z^d+A_{\rm IR}(\tau)]}.
\end{equation}
From Eq.\;(\ref{Eq:atto-carma_dipole_y_d}), there are two different longitudinal momentum $p_z^d$ for the same transverse momentum $p_y$, which is written as $p_z^d=\pm\sqrt{k_0^2-p_y^2}-A_{\rm IR}(\tau)=\pm p_{z0}-A_{\rm IR}(\tau)$. `+' and `-' correspond to the peak1 and peak2 as displayed in Fig.\;\ref{fig:figure2}, respectively. We use the notation $p_z^+$ and $p_z^-$ to mark these two peaks, i.e., $p_z^+=p_{z0}-A_{\rm IR}(\tau)$ and $p_z^-=-p_{z0}-A_{\rm IR}(\tau)$. Therefore, the momentum shifts for these two peaks are written as 
\begin{equation}{\label{Eq:Analysis_dpz}}
  \begin{aligned}
    \Delta p_z^+&=-p_y\frac{[2p_{z0}-A_{\rm IR}(\tau)]A_{\rm IR}(\tau)}{2cp_{z0}},\\
    \Delta p_z^-&=-p_y\frac{[2p_{z0}+A_{\rm IR}(\tau)]A_{\rm IR}(\tau)}{2cp_{z0}}.\\
  \end{aligned}
\end{equation}

\begin{figure}[t]
  \includegraphics[width=0.35\textwidth]{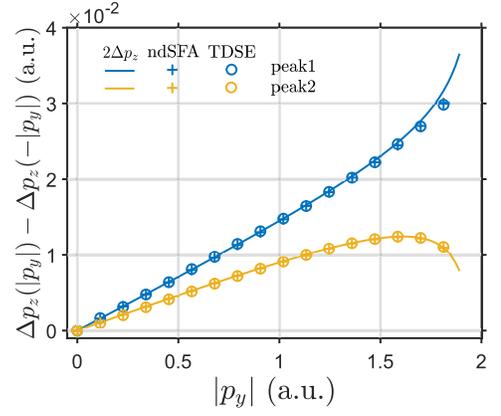}
  \caption{Longitudinal momentum shift between the PEMD of $|p_y|$ and $-|p_y|$ as a function of $|p_y|$. The solid lines are the results obtained by Eq.\;(\ref{Eq:Analysis_dpz}). The lines marked by `+' and `o' represent the results extracted from the PEMDs obtained by employing the ndSFA and numerically solving TDSE. The blue and yellow lines are the results for the peak1 and peak2 (shown in Fig.\;\ref{fig:figure2}), respectively.}
  \label{fig:figure5}
\end{figure}

To quantitatively demonstrate this momentum shift induced by the magnetic-field effect, we extract the peak position of the PEMD shown in Fig.\;\ref{fig:figure2} by fitting the momentum distribution with a Gaussian function. The peak shift between the PEMD for $|p_y|$ and $-|p_y|$ as a function of $|p_y|$ is displayed in Fig.\;\ref{fig:figure5} (marked by the `o'). We also calculate the PEMD by the ndSFA (Eq.\;(\ref{Eq:ndSFA:MP_d})), and the results extracted from the PEMD are also shown in Fig.\;\ref{fig:figure5} (marked by `+'). The solid lines are the results obtained by Eq.\;(\ref{Eq:Analysis_dpz}), i.e., $\Delta p_z(|p_y|)-\Delta p_z(-|p_y|)=2\Delta p_z$. Apparently, the results retrieved from the PEMDs agree remarkably with the analytical expression of the momentum shift. It indicates that the magnetic field of the streaking field induces an additional longitudinal momentum shift of the photoelectrons and this momentum shift strongly depends on the transverse photoelectrons momentum. We successfully reveal this momentum-dependent magnetic effect of the IR pulse by attosecond streaking. 

\begin{figure}[b]
  \includegraphics[width=0.45\textwidth]{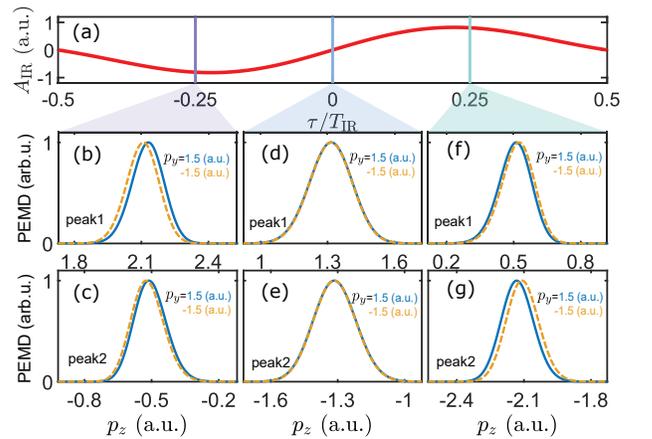}
  \caption{(a) Vector potential of the IR pulse. (b)-(g) Cuts of the PEMDs at $p_y=1.5$ (blue solid line) and $p_y=-1.5$ (yellow dashed line) with $\tau=-0.25T_{\rm IR}$(b, c), 0 (d, e) and $0.25T_{\rm IR}$ (f, g). The second and third rows are the PEMDs for peak1 and peak2, respectively.}
  \label{fig:figure6}
\end{figure}
\begin{figure}[t]
  \includegraphics[width=0.35\textwidth]{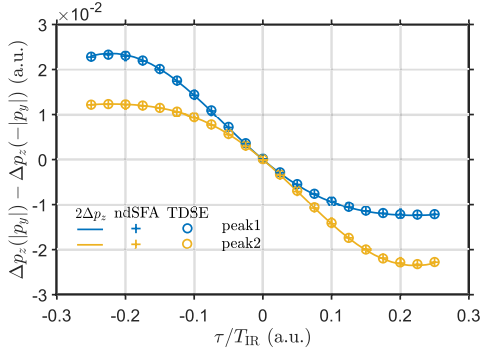}
  \caption{Longitudinal momentum shift between the PEMD of $|p_y|$ and $-|p_y|$ as a function of the time delay between the XUV and IR pulse. The color and marker of the lines are same as it in Fig.\;\ref{fig:figure5}.}
  \label{fig:figure7}
\end{figure}

Otherwise, as indicated by Eq.\;(\ref{Eq:Analysis_dpz}), the momentum shift not only depends on the transverse momentum of the photoelectrons, but also strongly depends on the vector potential of the IR pulse at the ionization time. To reveal this time-delay-dependent magnetic effect, the cuts of the PEMDs at $p_y=\pm1.5$ a.u. with various time delays are displayed in Fig.\;\ref{fig:figure6}. For $A_{\rm IR}(\tau)<0$, the momentum shift between the PEMD for $p_y>0$ and $p_y<0$ is towards the larger $p_z$, as displayed in Figs.\;\ref{fig:figure6}(b) and \ref{fig:figure6}(c). On the contrary, for $A_{\rm IR}(\tau)>0$, the momentum shift is towards the smaller $p_z$, as shown in Figs.\;\ref{fig:figure6}(f) and \ref{fig:figure6}(g). While for $A_{\rm IR}(\tau)=0$, the momentum shift vanishes, as indicated by Figs.\;\ref{fig:figure6}(d) and \ref{fig:figure6}(e). The behavior of the momentum shift qualitatively coincides with the prediction by Eq.\;(\ref{Eq:Analysis_dpz}). To quantitatively show the time-delay-dependent magnetic effect, we also extract the momentum shift from the PEMD at various time delays. The extracted procedure is the same as it in Fig.\;\ref{fig:figure5}. The results obtained from the different methods are shown in Fig.\;\ref{fig:figure7}. Apparently, the results retrieved from the PEMDs agree remarkably with the analytical expression of the momentum shift. It indicates that the time-resolved magnetic effect of the IR pulse is accurately revealed by the attosecond streaking technique. 

{ So far, we have successfully revealed the time-resolved magnetic-field effect in the laser-assisted XUV ionization. The magnetic field distorts the PEMD resulting in an  an time-delay- and $p_y$-dependent longitudinal momentum shift. This observable momentum shift can not be neglected in the interpretation of attosecond time-delay experiments.}

\subsection{Magnetic-field effect on the time shift}

\begin{figure}[t]
  \includegraphics[width=0.45\textwidth]{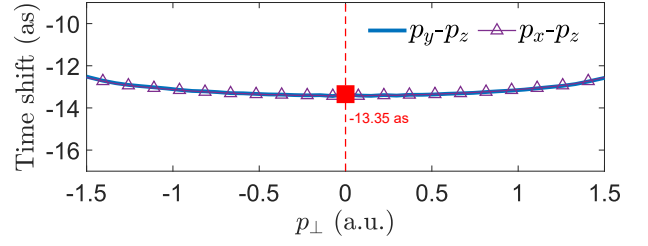}
  \caption{Time shift extracted from the PEMD obtained by numerically solving TDSE. The point marked by red square represents the results obtained from the PEMD with $p_x$=$p_y$=0 ($p_z>0$). The lines marked by the triangles and the blue solid line represent the results obtained from the PEMD in the $p_x$-$p_z$ ($p_y$=0) and $p_y$-$p_z$ ($p_x$=0) plane, respectively.}
  \label{fig:figure8}
\end{figure}

{As indicated by Eq.\;(\ref{Eq:Analysis_dpz}), there is no magnetic-field effect if the electrons are detected in the polarization direction, i.e., $p_y=0$. Thus, the time shift in the attosecond streaking discussed in previous works is not affected by the magnetic-field effect. To examine it, we extract the peak of the PEMD with $p_x=p_y=0$ ($p_z>0$) as a function of the time delay between the XUV and IR pulse by the Gaussian fitting. Then the time shift is obtained by fitting the changing of the peak position as $\tau$ with the shape of the $A_{\rm IR}$ field. The obtained time shift is $-13.35\pm1.1$ as, as displayed by the red square in Fig.\;\ref{fig:figure8}. As demonstrated in previous work \cite{Pazourek2015}, the time shift for the ionization of the ground state ($1s$) of H atom consists two parts
\begin{equation}\label{Eq:time_shift}
  \begin{aligned}
    t_s&=t_{\rm EWS}+t_{\rm CLC}\\
    &=\frac{d}{d\epsilon}\sigma_l^C(\epsilon)+\frac{1}{(2\epsilon)^{3/2}}[2-\ln(\epsilon T_{\rm IR})]. 
  \end{aligned}
\end{equation}
Here $t_{\rm EWS}$ represents the time shift induced by the short-range potential, which is the intrinsic atomic time delay. $t_{\rm CLC}$ is referred as the Coulomb-laser coupling (CLC) time shifts originating from the interaction between the long-range potential and the laser pulse. $\sigma_l^C(\epsilon)=\arg\Gamma(1+l-i/k)$. $\epsilon=\omega_{\rm XUV}-I_p$ and $k=\sqrt{2\epsilon}$. For the laser parameters considered in our work ($\lambda_{\rm IR}=3600$ nm, $\omega_{\rm XUV}=2.5$ a.u.), the time shift given by Eq.\;(\ref{Eq:time_shift}) is $t_s$=$-13.39$ as. Our numerical result agrees well with Eq.\;(\ref{Eq:time_shift}).

While for $p_y\ne0$, the magnetic-field effect results in an observable momentum shift $\Delta p_z^{\pm}$ in the order of 10$^{-2}$ a.u., i.e., $p_z(\tau)=p_{z0}-A_{\rm IR}(\tau)+\Delta p_z^{\pm}$. As indicated by previous works \cite{Ivanov2011,Pazourek2013,Pazourek2015}, $t_{\rm CLC}$ originates from the additional momentum shift induced by the CLC, i.e., $p_z(\tau)=p_{z0}-A_{\rm IR}(\tau)+\Delta p_z^{\rm CLC}$. The momentum shift is approximately given by $\Delta p_z^{\rm CLC}=t_{\rm CLC}E_{\rm IR}(\tau)$, where $E_{\rm IR}(t)=-dA_{\rm IR}(t)/dt$ is the electric field of the IR field. For the laser parameters considered in our work, this momentum shift is in the order of 10$^{-3}$ a.u.. At first glance, the momentum shift induced by the magnetic field will contribute an observable time shift as similar to the CLC. To examine it, the time shift as a function of the transverse momentum retrieved from the PEMD in the $p_y$-$p_z$ ($p_x$=0) plane is displayed by the blue solid line in Fig.\;\ref{fig:figure8}. The time shift extracted from the PEMD in the $p_x$-$p_z$ ($p_y$=0) plane is also shown by the line with the triangles for comparison, which can be taken as the reference without magnetic-field effect as indicated by Eq.\;(\ref{Eq:atto-carma_dipole}). No obvious difference between the results for the PEMDs in $p_x$-$p_z$ and $p_y$-$p_z$ plane is observed. It indicates that magnetic-field does not affect the time shift retrieved from the streaking spectrum. This is because the time-delay-dependent momentum shift induced by the magnetic-field effect is dependent on the terms of $A_{\rm IR}(\tau)$ and $A^2_{\rm IR}(\tau)$ as shown by Eq.\;(\ref{Eq:Analysis_dpz}). $A_{\rm}(\tau)$ does not contribute an additional time delay. While for $A_{\rm}(\tau)^2$, the frequency of this term equals to $2\omega_{\rm IR}$. And thus the contribution of this term is smeared out when fitting the time-delay-dependent peak of the PEMD by the $A_{\rm IR}$ field. Note that no magnetic-field effect on the time delay can not simply be attributed to that the magnetic-field effect is too small to be observed.
}

\section{conclusion}

In conclusion, the magnetic effect of the streaking field in laser-assisted XUV photoionization has been successfully revealed by numerically solving the TDSE and employing the ndSFA. We find that the magnetic effect results in a time-delay and transverse-momentum-dependent longitudinal momentum shift in the attosecond streaking camera. Although this momentum shift apparently alters the streaking spectrum, it does not affect the time shift retrieved from the streaking spectrum. Our results also indicate that the magnetic-field effect of the streaking field needs to be properly accounted for in a stronger and long-wavelength streaking field, such as the terahertz laser pulse, which has been used to investigate the time shift in atomic photoemission in the experiment \cite{Fruhling2009,Schmid2019}. 

Our work successfully reveals the magnetic-field effect in the attosecond streaking. It deepens our understanding of instantaneous magnetic-field effect on the photoelectrons and {paves way to investigate the time-resolved nondipole effect by attosecond technique {based on the laser-assisted XUV ionization, such as the reconstruction of attosecond bursts by beating of two-photon transitions (RABBITT)}. 

\section*{Acknowledgments}

We acknowledge helpful discussion with L. B. Madsen. This work is supported by National Key Research and Development Program of China (Grant No.2019YFA0308300) and National Natural Science Foundation of China (Grants No.11874163, No.12074265, and No.12021004). The computation is completed in the HPC Platform of Huazhong University of Science and Technology.

\appendix
\section{Nondipole Strong-field Approximation}
\label{Sec:appB}

With the dipole approximation, the transition amplitude of the photoelectrons ionized by the XUV pulse in the presence of the IR streaking field is written as \cite{Kitzler2002}
\begin{equation}
  M^{\rm SFA}_{\rm d}({\bf{p}},\tau)\approx-i\int_{0}^{\infty}dt\langle {\bf{p}}+{\bf A}_{\rm IR}(t)|{\bf E}_{\rm XUV}(t-\tau)\cdot{\bf{r}}|\psi_0\rangle e^{-iS_d({\bf{p}},t)}.
\end{equation}
Where ${\bf E}_{\rm XUV}(t)=-d{\bf A}_{\rm XUV}(t)/dt$ is the electric field of the XUV pulse. $\ket{\psi_0}$ represents the initial state of the electrons. The phase is expressed as $S_d({\bf{p}},t)=-I_pt+\int_t^{\infty}d\xi{[{\bf p}+{\bf A}(\xi)]^2}/2$. 
 
While with the nondipole corrections, the Volkov states can be expressed as (to the first order in 1/c),
\begin{equation}
  \ket{\psi_{\bf{p}}^V(t)}=e^{-iS({\bf{p}},t)}\ket{{\bf{p}}},
\end{equation}
consisting of plane waves $\ket{{\bf{p}}}$ and the generalized action \cite{Simon2021}
\begin{equation}
  S({\bf{p}},t)=\frac{1}{2}\int^td\xi[{\bf{p+A_{\rm IR}}}(\xi)+\frac{{\bf{e}}_y}{c}({\bf{p}}\cdot{\bf{A}}_{\rm IR}(\xi)+\frac{1}{2}{\bf{A}}_{\rm IR}^2(\xi))]^2.
\end{equation}
Inserting this nondipole Volkov states into the derivation of the transition amplitude with SFA, the nondipole effect modified transition amplitude can be straightforwardly written as
\begin{equation}{\label{Eq:ndSFA:MP}}
  M^{\rm SFA}_{\rm nd}({\bf{p}},\tau)\approx-i\int_{0}^{\infty}dt\langle {\bf{k}}(t)|{\bf E}_{\rm XUV}(t-\tau)\cdot{\bf{r}}|\psi_0\rangle e^{-iS_{\rm nd}({\bf{p}},t)}.
\end{equation}
Where $S_{\rm nd}({\bf{p}},t)=-I_pt+\int_t^{\infty}d\xi{{\bf{k}}^2(\xi)}/2$, and ${\bf{k}}(t)={\bf{p+A_{\rm IR}}}(t)+\frac{{\bf{e}}_y}{c}({\bf{p}}\cdot{\bf{A}}_{\rm IR}(t)+\frac{1}{2}{\bf{A}}_{\rm IR}^2(t))$. Since the XUV field includes the fast oscillations in time, it can be written as ${\bf E}_{\rm XUV}(t)=\widetilde{{\bf E}}_{\rm XUV}(t)e^{-i\omega_{\rm XUV}t}$. Here $\widetilde{{\bf E}}_{\rm XUV}(t)$ represents the complex valued envelope of the XUV pulse. Then, the transition amplitude is expressed as 
\begin{equation}
  M^{\rm SFA}_{\rm nd}({\bf{p}},\tau)\approx-i\int_{0}^{\infty}dt\langle {\bf{k}}(t)|\widetilde{\bf E}_{\rm XUV}(t-\tau)\cdot{\bf{r}}|\psi_0\rangle e^{-i(S_{\rm nd}({\bf{p}},t)+\omega_{\rm XUV}t)}.
\end{equation}
We mention that this transition amplitude only contains the nondipole effect of the IR field, and the nondipole effect of the XUV laser pulse is neglected. This is because the nondipole effect of the XUV pulse does not affect our results discussed in Sec.\;\ref{sec:mag_eff_IR} as indicated by Fig.\;\ref{fig:figure3}.

With the saddle point approximation, the nondipole modified attosecond streaking camera is obtained by
\begin{equation}
  \begin{aligned}
     \frac{\partial(S_{\rm nd}({\bf{p}},t)+\omega_{\rm XUV}t)}{\partial t}\bigg|_{t=t_i}&=0,\\
     \frac{1}{2}\left[{\bf{p+A_{\rm IR}}}(t_i)+\frac{{\bf{e}}_y}{c}\left({\bf{p}}\cdot{\bf{A}_{\rm IR}}(t_i)+\frac{1}{2}{\bf{A}}_{\rm IR}^2(t_i)\right)\right]^2&=\omega_{\rm XUV}-I_p.
  \end{aligned}
\end{equation}
In our calculation, the duration of the XUV pulse is extremely short. Therefore, the ionization time $t_i$ can be replaced by the time delay between the XUV and IR field, i.e., $t_i\approx\tau$. Then the attosecond streaking camera is expressed as 
\begin{equation}{\label{Eq:ndatto-camera}}
  \frac{1}{2}\left[{\bf{p+A_{\rm IR}}}(\tau)+\frac{{\bf{e}}_y}{c}\left({\bf{p}}\cdot{\bf{A}_{\rm IR}}(\tau)+\frac{1}{2}{\bf{A}}_{\rm IR}^2(\tau)\right)\right]^2=\omega_{\rm XUV}-I_p
\end{equation}

Let us first consider the PEMDs in the $p_x$-$p_z$ plane, i.e., $p_y=0$. In this case, Eq.\;\ref{Eq:ndatto-camera} is rewritten as 
\begin{equation}\label{Eq:atto-carma_dipole_py}
  \begin{aligned}
    &[p_z+A_{\rm IR}(\tau)]^2+p_x^2+\frac{1}{c^2}\left(p_zA_{\rm IR}(\tau)+\frac{1}{2}A_{\rm IR}^2(\tau)\right)^2\\
    =&2(\omega_{\rm XUV}-I_p)=k_0^2,
  \end{aligned}
\end{equation}
where $k_0$=$\sqrt{2(\omega_{\rm XUV}-I_p)}$. Within the order of $1/c$, this equation is reduced to 
\begin{equation}\label{Eq:atto-carma_dipole}
  [p_z+A_{\rm IR}(\tau)]^2+p_x^2=k_0^2.
\end{equation}
It is the same as it with the dipole approximation, which means that the PEMDs in the $p_x$-$p_z$ plane can be taken as a reference without nondipole corrections. 

While for the PEMDs in the $p_y$-$p_z$ plane, Eq.\;(\ref{Eq:ndatto-camera}) is written as 
\begin{equation}\label{Eq:atto-carma_nondipole_y}
  [p_z+A_{\rm IR}(\tau)]^2+\left[p_y+\frac{p_z}{c}A_{\rm IR}(\tau)+\frac{A_{\rm IR}^2(\tau)}{2c}\right]^2=k_0^2.
\end{equation}
With the dipole approximation, this relationship is given by
\begin{equation}\label{Eq:atto-carma_dipole_y}
  [p_z^d+A_{\rm IR}(\tau)]^2+p_y^2=k_0^2.
\end{equation}

\begin{thebibliography}{54}%
\makeatletter
\providecommand \@ifxundefined [1]{%
 \@ifx{#1\undefined}
}%
\providecommand \@ifnum [1]{%
 \ifnum #1\expandafter \@firstoftwo
 \else \expandafter \@secondoftwo
 \fi
}%
\providecommand \@ifx [1]{%
 \ifx #1\expandafter \@firstoftwo
 \else \expandafter \@secondoftwo
 \fi
}%
\providecommand \natexlab [1]{#1}%
\providecommand \enquote  [1]{``#1''}%
\providecommand \bibnamefont  [1]{#1}%
\providecommand \bibfnamefont [1]{#1}%
\providecommand \citenamefont [1]{#1}%
\providecommand \href@noop [0]{\@secondoftwo}%
\providecommand \href [0]{\begingroup \@sanitize@url \@href}%
\providecommand \@href[1]{\@@startlink{#1}\@@href}%
\providecommand \@@href[1]{\endgroup#1\@@endlink}%
\providecommand \@sanitize@url [0]{\catcode `\\12\catcode `\$12\catcode
  `\&12\catcode `\#12\catcode `\^12\catcode `\_12\catcode `\%12\relax}%
\providecommand \@@startlink[1]{}%
\providecommand \@@endlink[0]{}%
\providecommand \url  [0]{\begingroup\@sanitize@url \@url }%
\providecommand \@url [1]{\endgroup\@href {#1}{\urlprefix }}%
\providecommand \urlprefix  [0]{URL }%
\providecommand \Eprint [0]{\href }%
\providecommand \doibase [0]{http://dx.doi.org/}%
\providecommand \selectlanguage [0]{\@gobble}%
\providecommand \bibinfo  [0]{\@secondoftwo}%
\providecommand \bibfield  [0]{\@secondoftwo}%
\providecommand \translation [1]{[#1]}%
\providecommand \BibitemOpen [0]{}%
\providecommand \bibitemStop [0]{}%
\providecommand \bibitemNoStop [0]{.\EOS\space}%
\providecommand \EOS [0]{\spacefactor3000\relax}%
\providecommand \BibitemShut  [1]{\csname bibitem#1\endcsname}%
\let\auto@bib@innerbib\@empty
\bibitem [{\citenamefont {Milo{\v{s}}evi{\'{c}}}\ \emph
  {et~al.}(2006)\citenamefont {Milo{\v{s}}evi{\'{c}}}, \citenamefont {Paulus},
  \citenamefont {Bauer},\ and\ \citenamefont {Becker}}]{Milo_evi__2006}%
  \BibitemOpen
  \bibfield  {author} {\bibinfo {author} {\bibfnamefont {D.~B.}\ \bibnamefont
  {Milo{\v{s}}evi{\'{c}}}}, \bibinfo {author} {\bibfnamefont {G.~G.}\
  \bibnamefont {Paulus}}, \bibinfo {author} {\bibfnamefont {D.}~\bibnamefont
  {Bauer}}, \ and\ \bibinfo {author} {\bibfnamefont {W.}~\bibnamefont
  {Becker}},\ }\bibfield  {title} {\enquote {\bibinfo {title} {Above-threshold
  ionization by few-cycle pulses},}\ }\href {\doibase
  10.1088/0953-4075/39/14/r01} {\bibfield  {journal} {\bibinfo  {journal} {J.
  Phys. B}\ }\textbf {\bibinfo {volume} {39}},\ \bibinfo {pages} {R203}
  (\bibinfo {year} {2006})}\BibitemShut {NoStop}%
\bibitem [{\citenamefont {Reiss}(2008)}]{Reiss2008}%
  \BibitemOpen
  \bibfield  {author} {\bibinfo {author} {\bibfnamefont {H.~R.}\ \bibnamefont
  {Reiss}},\ }\bibfield  {title} {\enquote {\bibinfo {title} {Limits on
  tunneling theories of strong-field ionization},}\ }\href {\doibase
  10.1103/PhysRevLett.101.043002} {\bibfield  {journal} {\bibinfo  {journal}
  {Phys. Rev. Lett.}\ }\textbf {\bibinfo {volume} {101}},\ \bibinfo {pages}
  {043002} (\bibinfo {year} {2008})}\BibitemShut {NoStop}%
\bibitem [{\citenamefont {Smeenk}\ \emph {et~al.}(2011)\citenamefont {Smeenk},
  \citenamefont {Arissian}, \citenamefont {Zhou}, \citenamefont {Mysyrowicz},
  \citenamefont {Villeneuve}, \citenamefont {Staudte},\ and\ \citenamefont
  {Corkum}}]{Smeenk2011}%
  \BibitemOpen
  \bibfield  {author} {\bibinfo {author} {\bibfnamefont {C.~T.~L.}\
  \bibnamefont {Smeenk}}, \bibinfo {author} {\bibfnamefont {L.}~\bibnamefont
  {Arissian}}, \bibinfo {author} {\bibfnamefont {B.}~\bibnamefont {Zhou}},
  \bibinfo {author} {\bibfnamefont {A.}~\bibnamefont {Mysyrowicz}}, \bibinfo
  {author} {\bibfnamefont {D.~M.}\ \bibnamefont {Villeneuve}}, \bibinfo
  {author} {\bibfnamefont {A.}~\bibnamefont {Staudte}}, \ and\ \bibinfo
  {author} {\bibfnamefont {P.~B.}\ \bibnamefont {Corkum}},\ }\bibfield  {title}
  {\enquote {\bibinfo {title} {Partitioning of the linear photon momentum in
  multiphoton ionization},}\ }\href {\doibase 10.1103/PhysRevLett.106.193002}
  {\bibfield  {journal} {\bibinfo  {journal} {Phys. Rev. Lett.}\ }\textbf
  {\bibinfo {volume} {106}},\ \bibinfo {pages} {193002} (\bibinfo {year}
  {2011})}\BibitemShut {NoStop}%
\bibitem [{\citenamefont {Ludwig}\ \emph {et~al.}(2014)\citenamefont {Ludwig},
  \citenamefont {Maurer}, \citenamefont {Mayer}, \citenamefont {Phillips},
  \citenamefont {Gallmann},\ and\ \citenamefont {Keller}}]{Ludwig2014}%
  \BibitemOpen
  \bibfield  {author} {\bibinfo {author} {\bibfnamefont {A.}~\bibnamefont
  {Ludwig}}, \bibinfo {author} {\bibfnamefont {J.}~\bibnamefont {Maurer}},
  \bibinfo {author} {\bibfnamefont {B.~W.}\ \bibnamefont {Mayer}}, \bibinfo
  {author} {\bibfnamefont {C.~R.}\ \bibnamefont {Phillips}}, \bibinfo {author}
  {\bibfnamefont {L.}~\bibnamefont {Gallmann}}, \ and\ \bibinfo {author}
  {\bibfnamefont {U.}~\bibnamefont {Keller}},\ }\bibfield  {title} {\enquote
  {\bibinfo {title} {Breakdown of the dipole approximation in strong-field
  ionization},}\ }\href {\doibase 10.1103/PhysRevLett.113.243001} {\bibfield
  {journal} {\bibinfo  {journal} {Phys. Rev. Lett.}\ }\textbf {\bibinfo
  {volume} {113}},\ \bibinfo {pages} {243001} (\bibinfo {year}
  {2014})}\BibitemShut {NoStop}%
\bibitem [{\citenamefont {Hartung}\ \emph {et~al.}(2019)\citenamefont
  {Hartung}, \citenamefont {Eckart}, \citenamefont {Brennecke}, \citenamefont
  {Rist}, \citenamefont {Trabert}, \citenamefont {Fehre}, \citenamefont
  {Richter}, \citenamefont {Sann}, \citenamefont {Zeller}, \citenamefont
  {Henrichs}, \citenamefont {Kastirke}, \citenamefont {Hoehl}, \citenamefont
  {Kalinin}, \citenamefont {Sch{\"{o}}ffler}, \citenamefont {Jahnke},
  \citenamefont {Schmidt}, \citenamefont {Lein}, \citenamefont {Kunitski},\
  and\ \citenamefont {D{\"{o}}rner}}]{Hartung2019}%
  \BibitemOpen
  \bibfield  {author} {\bibinfo {author} {\bibfnamefont {A.}~\bibnamefont
  {Hartung}}, \bibinfo {author} {\bibfnamefont {S.}~\bibnamefont {Eckart}},
  \bibinfo {author} {\bibfnamefont {S.}~\bibnamefont {Brennecke}}, \bibinfo
  {author} {\bibfnamefont {J.}~\bibnamefont {Rist}}, \bibinfo {author}
  {\bibfnamefont {D.}~\bibnamefont {Trabert}}, \bibinfo {author} {\bibfnamefont
  {K.}~\bibnamefont {Fehre}}, \bibinfo {author} {\bibfnamefont
  {M.}~\bibnamefont {Richter}}, \bibinfo {author} {\bibfnamefont
  {H.}~\bibnamefont {Sann}}, \bibinfo {author} {\bibfnamefont {S.}~\bibnamefont
  {Zeller}}, \bibinfo {author} {\bibfnamefont {K.}~\bibnamefont {Henrichs}},
  \bibinfo {author} {\bibfnamefont {G.}~\bibnamefont {Kastirke}}, \bibinfo
  {author} {\bibfnamefont {J.}~\bibnamefont {Hoehl}}, \bibinfo {author}
  {\bibfnamefont {A.}~\bibnamefont {Kalinin}}, \bibinfo {author} {\bibfnamefont
  {M.~S.}\ \bibnamefont {Sch{\"{o}}ffler}}, \bibinfo {author} {\bibfnamefont
  {T.}~\bibnamefont {Jahnke}}, \bibinfo {author} {\bibfnamefont {L.~Ph~H.}\
  \bibnamefont {Schmidt}}, \bibinfo {author} {\bibfnamefont {M.}~\bibnamefont
  {Lein}}, \bibinfo {author} {\bibfnamefont {M.}~\bibnamefont {Kunitski}}, \
  and\ \bibinfo {author} {\bibfnamefont {R.}~\bibnamefont {D{\"{o}}rner}},\
  }\bibfield  {title} {\enquote {\bibinfo {title} {{Magnetic fields alter
  strong-field ionization}},}\ }\href {\doibase 10.1038/s41567-019-0653-y}
  {\bibfield  {journal} {\bibinfo  {journal} {Nat. Phys.}\ }\textbf {\bibinfo
  {volume} {15}},\ \bibinfo {pages} {1222} (\bibinfo {year}
  {2019})}\BibitemShut {NoStop}%
\bibitem [{\citenamefont {Lin}\ \emph {et~al.}(2022{\natexlab{a}})\citenamefont
  {Lin}, \citenamefont {Brennecke}, \citenamefont {Ni}, \citenamefont {Chen},
  \citenamefont {Hartung}, \citenamefont {Trabert}, \citenamefont {Fehre},
  \citenamefont {Rist}, \citenamefont {Tong}, \citenamefont {Burgd\"orfer},
  \citenamefont {Schmidt}, \citenamefont {Sch\"offler}, \citenamefont {Jahnke},
  \citenamefont {Kunitski}, \citenamefont {He}, \citenamefont {Lein},
  \citenamefont {Eckart},\ and\ \citenamefont {D\"orner}}]{Lin2022}%
  \BibitemOpen
  \bibfield  {author} {\bibinfo {author} {\bibfnamefont {K.}~\bibnamefont
  {Lin}}, \bibinfo {author} {\bibfnamefont {S.}~\bibnamefont {Brennecke}},
  \bibinfo {author} {\bibfnamefont {H.}~\bibnamefont {Ni}}, \bibinfo {author}
  {\bibfnamefont {X.}~\bibnamefont {Chen}}, \bibinfo {author} {\bibfnamefont
  {A.}~\bibnamefont {Hartung}}, \bibinfo {author} {\bibfnamefont
  {D.}~\bibnamefont {Trabert}}, \bibinfo {author} {\bibfnamefont
  {K.}~\bibnamefont {Fehre}}, \bibinfo {author} {\bibfnamefont
  {J.}~\bibnamefont {Rist}}, \bibinfo {author} {\bibfnamefont {X.~M.}\
  \bibnamefont {Tong}}, \bibinfo {author} {\bibfnamefont {J.}~\bibnamefont
  {Burgd\"orfer}}, \bibinfo {author} {\bibfnamefont {L.~Ph.~H.}\ \bibnamefont
  {Schmidt}}, \bibinfo {author} {\bibfnamefont {M.~S.}\ \bibnamefont
  {Sch\"offler}}, \bibinfo {author} {\bibfnamefont {T.}~\bibnamefont {Jahnke}},
  \bibinfo {author} {\bibfnamefont {M.}~\bibnamefont {Kunitski}}, \bibinfo
  {author} {\bibfnamefont {F.}~\bibnamefont {He}}, \bibinfo {author}
  {\bibfnamefont {M.}~\bibnamefont {Lein}}, \bibinfo {author} {\bibfnamefont
  {S.}~\bibnamefont {Eckart}}, \ and\ \bibinfo {author} {\bibfnamefont
  {R.}~\bibnamefont {D\"orner}},\ }\bibfield  {title} {\enquote {\bibinfo
  {title} {Magnetic-field effect in high-order above-threshold ionization},}\
  }\href {\doibase 10.1103/PhysRevLett.128.023201} {\bibfield  {journal}
  {\bibinfo  {journal} {Phys. Rev. Lett.}\ }\textbf {\bibinfo {volume} {128}},\
  \bibinfo {pages} {023201} (\bibinfo {year} {2022}{\natexlab{a}})}\BibitemShut
  {NoStop}%
\bibitem [{\citenamefont {Lin}\ \emph {et~al.}(2022{\natexlab{b}})\citenamefont
  {Lin}, \citenamefont {Chen}, \citenamefont {Eckart}, \citenamefont {Jiang},
  \citenamefont {Hartung}, \citenamefont {Trabert}, \citenamefont {Fehre},
  \citenamefont {Rist}, \citenamefont {Schmidt}, \citenamefont {Sch\"offler},
  \citenamefont {Jahnke}, \citenamefont {Kunitski}, \citenamefont {He},\ and\
  \citenamefont {D\"orner}}]{Lin2022_1}%
  \BibitemOpen
  \bibfield  {author} {\bibinfo {author} {\bibfnamefont {K.}~\bibnamefont
  {Lin}}, \bibinfo {author} {\bibfnamefont {X.}~\bibnamefont {Chen}}, \bibinfo
  {author} {\bibfnamefont {S.}~\bibnamefont {Eckart}}, \bibinfo {author}
  {\bibfnamefont {H.}~\bibnamefont {Jiang}}, \bibinfo {author} {\bibfnamefont
  {A.}~\bibnamefont {Hartung}}, \bibinfo {author} {\bibfnamefont
  {D.}~\bibnamefont {Trabert}}, \bibinfo {author} {\bibfnamefont
  {K.}~\bibnamefont {Fehre}}, \bibinfo {author} {\bibfnamefont
  {J.}~\bibnamefont {Rist}}, \bibinfo {author} {\bibfnamefont {L.~Ph.~H.}\
  \bibnamefont {Schmidt}}, \bibinfo {author} {\bibfnamefont {M.~S.}\
  \bibnamefont {Sch\"offler}}, \bibinfo {author} {\bibfnamefont
  {T.}~\bibnamefont {Jahnke}}, \bibinfo {author} {\bibfnamefont
  {M.}~\bibnamefont {Kunitski}}, \bibinfo {author} {\bibfnamefont
  {F.}~\bibnamefont {He}}, \ and\ \bibinfo {author} {\bibfnamefont
  {R.}~\bibnamefont {D\"orner}},\ }\bibfield  {title} {\enquote {\bibinfo
  {title} {Magnetic-field effect as a tool to investigate electron correlation
  in strong-field ionization},}\ }\href {\doibase
  10.1103/PhysRevLett.128.113201} {\bibfield  {journal} {\bibinfo  {journal}
  {Phys. Rev. Lett.}\ }\textbf {\bibinfo {volume} {128}},\ \bibinfo {pages}
  {113201} (\bibinfo {year} {2022}{\natexlab{b}})}\BibitemShut {NoStop}%
\bibitem [{\citenamefont {Sun}\ \emph {et~al.}(2020)\citenamefont {Sun},
  \citenamefont {Chen}, \citenamefont {Zhang}, \citenamefont {Qiang},
  \citenamefont {Li}, \citenamefont {Lu}, \citenamefont {Gong}, \citenamefont
  {Ji}, \citenamefont {Lin}, \citenamefont {Li}, \citenamefont {Tong},
  \citenamefont {Chen}, \citenamefont {Ruiz}, \citenamefont {Wu},\ and\
  \citenamefont {He}}]{Sun2020}%
  \BibitemOpen
  \bibfield  {author} {\bibinfo {author} {\bibfnamefont {F.}~\bibnamefont
  {Sun}}, \bibinfo {author} {\bibfnamefont {X.}~\bibnamefont {Chen}}, \bibinfo
  {author} {\bibfnamefont {W.}~\bibnamefont {Zhang}}, \bibinfo {author}
  {\bibfnamefont {J.}~\bibnamefont {Qiang}}, \bibinfo {author} {\bibfnamefont
  {H.}~\bibnamefont {Li}}, \bibinfo {author} {\bibfnamefont {P.}~\bibnamefont
  {Lu}}, \bibinfo {author} {\bibfnamefont {X.}~\bibnamefont {Gong}}, \bibinfo
  {author} {\bibfnamefont {Q.}~\bibnamefont {Ji}}, \bibinfo {author}
  {\bibfnamefont {K.}~\bibnamefont {Lin}}, \bibinfo {author} {\bibfnamefont
  {H.}~\bibnamefont {Li}}, \bibinfo {author} {\bibfnamefont {J.}~\bibnamefont
  {Tong}}, \bibinfo {author} {\bibfnamefont {F.}~\bibnamefont {Chen}}, \bibinfo
  {author} {\bibfnamefont {C.}~\bibnamefont {Ruiz}}, \bibinfo {author}
  {\bibfnamefont {J.}~\bibnamefont {Wu}}, \ and\ \bibinfo {author}
  {\bibfnamefont {F.}~\bibnamefont {He}},\ }\bibfield  {title} {\enquote
  {\bibinfo {title} {Longitudinal photon-momentum transfer in strong-field
  double ionization of argon atoms},}\ }\href {\doibase
  10.1103/PhysRevA.101.021402} {\bibfield  {journal} {\bibinfo  {journal}
  {Phys. Rev. A}\ }\textbf {\bibinfo {volume} {101}},\ \bibinfo {pages}
  {021402(R)} (\bibinfo {year} {2020})}\BibitemShut {NoStop}%
\bibitem [{\citenamefont {Chen}\ \emph {et~al.}(2020)\citenamefont {Chen},
  \citenamefont {Jiang}, \citenamefont {Grundmann}, \citenamefont {Trinter},
  \citenamefont {Sch\"offler}, \citenamefont {Jahnke}, \citenamefont
  {D\"orner}, \citenamefont {Liang}, \citenamefont {Wang}, \citenamefont
  {Peng},\ and\ \citenamefont {Gong}}]{Chen2020}%
  \BibitemOpen
  \bibfield  {author} {\bibinfo {author} {\bibfnamefont {S.~G.}\ \bibnamefont
  {Chen}}, \bibinfo {author} {\bibfnamefont {W.~C.}\ \bibnamefont {Jiang}},
  \bibinfo {author} {\bibfnamefont {S.}~\bibnamefont {Grundmann}}, \bibinfo
  {author} {\bibfnamefont {F.}~\bibnamefont {Trinter}}, \bibinfo {author}
  {\bibfnamefont {M.~S.}\ \bibnamefont {Sch\"offler}}, \bibinfo {author}
  {\bibfnamefont {T.}~\bibnamefont {Jahnke}}, \bibinfo {author} {\bibfnamefont
  {R.}~\bibnamefont {D\"orner}}, \bibinfo {author} {\bibfnamefont
  {H.}~\bibnamefont {Liang}}, \bibinfo {author} {\bibfnamefont {M.~X.}\
  \bibnamefont {Wang}}, \bibinfo {author} {\bibfnamefont {L.~Y.}\ \bibnamefont
  {Peng}}, \ and\ \bibinfo {author} {\bibfnamefont {Q.}~\bibnamefont {Gong}},\
  }\bibfield  {title} {\enquote {\bibinfo {title} {Photon momentum transfer in
  single-photon double ionization of helium},}\ }\href {\doibase
  10.1103/PhysRevLett.124.043201} {\bibfield  {journal} {\bibinfo  {journal}
  {Phys. Rev. Lett.}\ }\textbf {\bibinfo {volume} {124}},\ \bibinfo {pages}
  {043201} (\bibinfo {year} {2020})}\BibitemShut {NoStop}%
\bibitem [{\citenamefont {Grundmann}\ \emph {et~al.}(2020)\citenamefont
  {Grundmann}, \citenamefont {Kircher}, \citenamefont {Vela-Perez},
  \citenamefont {Nalin}, \citenamefont {Trabert}, \citenamefont {Anders},
  \citenamefont {Melzer}, \citenamefont {Rist}, \citenamefont {Pier},
  \citenamefont {Strenger}, \citenamefont {Siebert}, \citenamefont {Demekhin},
  \citenamefont {Schmidt}, \citenamefont {Trinter}, \citenamefont
  {Sch\"offler}, \citenamefont {Jahnke},\ and\ \citenamefont
  {D\"orner}}]{Grundmann2020}%
  \BibitemOpen
  \bibfield  {author} {\bibinfo {author} {\bibfnamefont {S.}~\bibnamefont
  {Grundmann}}, \bibinfo {author} {\bibfnamefont {M.}~\bibnamefont {Kircher}},
  \bibinfo {author} {\bibfnamefont {I.}~\bibnamefont {Vela-Perez}}, \bibinfo
  {author} {\bibfnamefont {G.}~\bibnamefont {Nalin}}, \bibinfo {author}
  {\bibfnamefont {D.}~\bibnamefont {Trabert}}, \bibinfo {author} {\bibfnamefont
  {N.}~\bibnamefont {Anders}}, \bibinfo {author} {\bibfnamefont
  {N.}~\bibnamefont {Melzer}}, \bibinfo {author} {\bibfnamefont
  {J.}~\bibnamefont {Rist}}, \bibinfo {author} {\bibfnamefont {A.}~\bibnamefont
  {Pier}}, \bibinfo {author} {\bibfnamefont {N.}~\bibnamefont {Strenger}},
  \bibinfo {author} {\bibfnamefont {J.}~\bibnamefont {Siebert}}, \bibinfo
  {author} {\bibfnamefont {P.~V.}\ \bibnamefont {Demekhin}}, \bibinfo {author}
  {\bibfnamefont {L.~Ph.~H.}\ \bibnamefont {Schmidt}}, \bibinfo {author}
  {\bibfnamefont {F.}~\bibnamefont {Trinter}}, \bibinfo {author} {\bibfnamefont
  {M.~S.}\ \bibnamefont {Sch\"offler}}, \bibinfo {author} {\bibfnamefont
  {T.}~\bibnamefont {Jahnke}}, \ and\ \bibinfo {author} {\bibfnamefont
  {R.}~\bibnamefont {D\"orner}},\ }\bibfield  {title} {\enquote {\bibinfo
  {title} {Observation of photoion backward emission in photoionization of he
  and ${\mathrm{n}}_{2}$},}\ }\href {\doibase 10.1103/PhysRevLett.124.233201}
  {\bibfield  {journal} {\bibinfo  {journal} {Phys. Rev. Lett.}\ }\textbf
  {\bibinfo {volume} {124}},\ \bibinfo {pages} {233201} (\bibinfo {year}
  {2020})}\BibitemShut {NoStop}%
\bibitem [{\citenamefont {Klaiber}\ \emph {et~al.}(2013)\citenamefont
  {Klaiber}, \citenamefont {Yakaboylu}, \citenamefont {Bauke}, \citenamefont
  {Hatsagortsyan},\ and\ \citenamefont {Keitel}}]{Klaiber2013}%
  \BibitemOpen
  \bibfield  {author} {\bibinfo {author} {\bibfnamefont {M.}~\bibnamefont
  {Klaiber}}, \bibinfo {author} {\bibfnamefont {E.}~\bibnamefont {Yakaboylu}},
  \bibinfo {author} {\bibfnamefont {H.}~\bibnamefont {Bauke}}, \bibinfo
  {author} {\bibfnamefont {K.~Z.}\ \bibnamefont {Hatsagortsyan}}, \ and\
  \bibinfo {author} {\bibfnamefont {C.~H.}\ \bibnamefont {Keitel}},\ }\bibfield
   {title} {\enquote {\bibinfo {title} {Under-the-barrier dynamics in
  laser-induced relativistic tunneling},}\ }\href {\doibase
  10.1103/PhysRevLett.110.153004} {\bibfield  {journal} {\bibinfo  {journal}
  {Phys. Rev. Lett.}\ }\textbf {\bibinfo {volume} {110}},\ \bibinfo {pages}
  {153004} (\bibinfo {year} {2013})}\BibitemShut {NoStop}%
\bibitem [{\citenamefont {Titi}\ and\ \citenamefont {Drake}(2012)}]{Titi2012}%
  \BibitemOpen
  \bibfield  {author} {\bibinfo {author} {\bibfnamefont {A.~S.}\ \bibnamefont
  {Titi}}\ and\ \bibinfo {author} {\bibfnamefont {G.~W.~F.}\ \bibnamefont
  {Drake}},\ }\bibfield  {title} {\enquote {\bibinfo {title} {Quantum theory of
  longitudinal momentum transfer in above-threshold ionization},}\ }\href
  {\doibase 10.1103/PhysRevA.85.041404} {\bibfield  {journal} {\bibinfo
  {journal} {Phys. Rev. A}\ }\textbf {\bibinfo {volume} {85}},\ \bibinfo
  {pages} {041404(R)} (\bibinfo {year} {2012})}\BibitemShut {NoStop}%
\bibitem [{\citenamefont {Jensen}\ \emph {et~al.}(2020)\citenamefont {Jensen},
  \citenamefont {Lund},\ and\ \citenamefont {Madsen}}]{Jensen2020}%
  \BibitemOpen
  \bibfield  {author} {\bibinfo {author} {\bibfnamefont {S.~V.~B.}\
  \bibnamefont {Jensen}}, \bibinfo {author} {\bibfnamefont {M.~M.}\
  \bibnamefont {Lund}}, \ and\ \bibinfo {author} {\bibfnamefont {L.~B.}\
  \bibnamefont {Madsen}},\ }\bibfield  {title} {\enquote {\bibinfo {title}
  {Nondipole strong-field-approximation hamiltonian},}\ }\href {\doibase
  10.1103/PhysRevA.101.043408} {\bibfield  {journal} {\bibinfo  {journal}
  {Phys. Rev. A}\ }\textbf {\bibinfo {volume} {101}},\ \bibinfo {pages}
  {043408} (\bibinfo {year} {2020})}\BibitemShut {NoStop}%
\bibitem [{\citenamefont {Chelkowski}\ \emph {et~al.}(2014)\citenamefont
  {Chelkowski}, \citenamefont {Bandrauk},\ and\ \citenamefont
  {Corkum}}]{Chelkowski2014}%
  \BibitemOpen
  \bibfield  {author} {\bibinfo {author} {\bibfnamefont {S.}~\bibnamefont
  {Chelkowski}}, \bibinfo {author} {\bibfnamefont {A.~D.}\ \bibnamefont
  {Bandrauk}}, \ and\ \bibinfo {author} {\bibfnamefont {P.~B.}\ \bibnamefont
  {Corkum}},\ }\bibfield  {title} {\enquote {\bibinfo {title} {Photon momentum
  sharing between an electron and an ion in photoionization: From one-photon
  (photoelectric effect) to multiphoton absorption},}\ }\href {\doibase
  10.1103/PhysRevLett.113.263005} {\bibfield  {journal} {\bibinfo  {journal}
  {Phys. Rev. Lett.}\ }\textbf {\bibinfo {volume} {113}},\ \bibinfo {pages}
  {263005} (\bibinfo {year} {2014})}\BibitemShut {NoStop}%
\bibitem [{\citenamefont {Chelkowski}\ \emph {et~al.}(2017)\citenamefont
  {Chelkowski}, \citenamefont {Bandrauk},\ and\ \citenamefont
  {Corkum}}]{Chelkowski2017}%
  \BibitemOpen
  \bibfield  {author} {\bibinfo {author} {\bibfnamefont {S.}~\bibnamefont
  {Chelkowski}}, \bibinfo {author} {\bibfnamefont {A.~D.}\ \bibnamefont
  {Bandrauk}}, \ and\ \bibinfo {author} {\bibfnamefont {P.~B.}\ \bibnamefont
  {Corkum}},\ }\bibfield  {title} {\enquote {\bibinfo {title} {Photon-momentum
  transfer in photoionization: From few photons to many},}\ }\href {\doibase
  10.1103/PhysRevA.95.053402} {\bibfield  {journal} {\bibinfo  {journal} {Phys.
  Rev. A}\ }\textbf {\bibinfo {volume} {95}},\ \bibinfo {pages} {053402}
  (\bibinfo {year} {2017})}\BibitemShut {NoStop}%
\bibitem [{\citenamefont {Chelkowski}\ \emph {et~al.}(2015)\citenamefont
  {Chelkowski}, \citenamefont {Bandrauk},\ and\ \citenamefont
  {Corkum}}]{Chelkowski2015}%
  \BibitemOpen
  \bibfield  {author} {\bibinfo {author} {\bibfnamefont {S.}~\bibnamefont
  {Chelkowski}}, \bibinfo {author} {\bibfnamefont {A.~D.}\ \bibnamefont
  {Bandrauk}}, \ and\ \bibinfo {author} {\bibfnamefont {P.~B.}\ \bibnamefont
  {Corkum}},\ }\bibfield  {title} {\enquote {\bibinfo {title} {Photon-momentum
  transfer in multiphoton ionization and in time-resolved holography with
  photoelectrons},}\ }\href {\doibase 10.1103/PhysRevA.92.051401} {\bibfield
  {journal} {\bibinfo  {journal} {Phys. Rev. A}\ }\textbf {\bibinfo {volume}
  {92}},\ \bibinfo {pages} {051401(R)} (\bibinfo {year} {2015})}\BibitemShut
  {NoStop}%
\bibitem [{\citenamefont {He}\ \emph {et~al.}(2017)\citenamefont {He},
  \citenamefont {Lao},\ and\ \citenamefont {He}}]{He2017}%
  \BibitemOpen
  \bibfield  {author} {\bibinfo {author} {\bibfnamefont {P.~L.}\ \bibnamefont
  {He}}, \bibinfo {author} {\bibfnamefont {D.}~\bibnamefont {Lao}}, \ and\
  \bibinfo {author} {\bibfnamefont {F.}~\bibnamefont {He}},\ }\bibfield
  {title} {\enquote {\bibinfo {title} {Strong field theories beyond dipole
  approximations in nonrelativistic regimes},}\ }\href {\doibase
  10.1103/PhysRevLett.118.163203} {\bibfield  {journal} {\bibinfo  {journal}
  {Phys. Rev. Lett.}\ }\textbf {\bibinfo {volume} {118}},\ \bibinfo {pages}
  {163203} (\bibinfo {year} {2017})}\BibitemShut {NoStop}%
\bibitem [{\citenamefont {Liu}\ \emph {et~al.}(2013)\citenamefont {Liu},
  \citenamefont {Xia}, \citenamefont {Tao},\ and\ \citenamefont
  {Fu}}]{Liu2013}%
  \BibitemOpen
  \bibfield  {author} {\bibinfo {author} {\bibfnamefont {J.}~\bibnamefont
  {Liu}}, \bibinfo {author} {\bibfnamefont {Q.~Z.}\ \bibnamefont {Xia}},
  \bibinfo {author} {\bibfnamefont {J.~F.}\ \bibnamefont {Tao}}, \ and\
  \bibinfo {author} {\bibfnamefont {L.~B.}\ \bibnamefont {Fu}},\ }\bibfield
  {title} {\enquote {\bibinfo {title} {Coulomb effects in photon-momentum
  partitioning during atomic ionization by intense linearly polarized light},}\
  }\href {\doibase 10.1103/PhysRevA.87.041403} {\bibfield  {journal} {\bibinfo
  {journal} {Phys. Rev. A}\ }\textbf {\bibinfo {volume} {87}},\ \bibinfo
  {pages} {041403(R)} (\bibinfo {year} {2013})}\BibitemShut {NoStop}%
\bibitem [{\citenamefont {Wang}\ \emph {et~al.}(2017)\citenamefont {Wang},
  \citenamefont {Xiao}, \citenamefont {Liang}, \citenamefont {Chen},\ and\
  \citenamefont {Peng}}]{Wang2017}%
  \BibitemOpen
  \bibfield  {author} {\bibinfo {author} {\bibfnamefont {M.~X.}\ \bibnamefont
  {Wang}}, \bibinfo {author} {\bibfnamefont {X.~R.}\ \bibnamefont {Xiao}},
  \bibinfo {author} {\bibfnamefont {H.}~\bibnamefont {Liang}}, \bibinfo
  {author} {\bibfnamefont {S.~G.}\ \bibnamefont {Chen}}, \ and\ \bibinfo
  {author} {\bibfnamefont {L.~Y.}\ \bibnamefont {Peng}},\ }\bibfield  {title}
  {\enquote {\bibinfo {title} {{Photon-momentum transfer in one- and two-photon
  ionization of atoms}},}\ }\href {\doibase 10.1103/PhysRevA.96.043414}
  {\bibfield  {journal} {\bibinfo  {journal} {Phys. Rev. A}\ }\textbf {\bibinfo
  {volume} {96}},\ \bibinfo {pages} {043414} (\bibinfo {year}
  {2017})}\BibitemShut {NoStop}%
\bibitem [{\citenamefont {Maurer}\ and\ \citenamefont
  {Keller}(2021)}]{Maurer2021}%
  \BibitemOpen
  \bibfield  {author} {\bibinfo {author} {\bibfnamefont {J.}~\bibnamefont
  {Maurer}}\ and\ \bibinfo {author} {\bibfnamefont {U.}~\bibnamefont
  {Keller}},\ }\bibfield  {title} {\enquote {\bibinfo {title} {{Ionization in
  intense laser fields beyond the electric dipole approximation: concepts,
  methods, achievements and future directions}},}\ }\href {\doibase
  10.1088/1361-6455/abf731} {\bibfield  {journal} {\bibinfo  {journal} {J.
  Phys. B}\ }\textbf {\bibinfo {volume} {54}},\ \bibinfo {pages} {094001}
  (\bibinfo {year} {2021})}\BibitemShut {NoStop}%
\bibitem [{\citenamefont {Willenberg}\ \emph {et~al.}(2019)\citenamefont
  {Willenberg}, \citenamefont {Maurer}, \citenamefont {Mayer},\ and\
  \citenamefont {Keller}}]{Willenberg2019}%
  \BibitemOpen
  \bibfield  {author} {\bibinfo {author} {\bibfnamefont {B.}~\bibnamefont
  {Willenberg}}, \bibinfo {author} {\bibfnamefont {J.}~\bibnamefont {Maurer}},
  \bibinfo {author} {\bibfnamefont {B.~W.}\ \bibnamefont {Mayer}}, \ and\
  \bibinfo {author} {\bibfnamefont {U.}~\bibnamefont {Keller}},\ }\bibfield
  {title} {\enquote {\bibinfo {title} {{Sub-cycle time resolution of
  multi-photon momentum transfer in strong-field ionization}},}\ }\href
  {\doibase 10.1038/s41467-019-13409-6} {\bibfield  {journal} {\bibinfo
  {journal} {Nat. Commun.}\ }\textbf {\bibinfo {volume} {10}},\ \bibinfo
  {pages} {5548} (\bibinfo {year} {2019})}\BibitemShut {NoStop}%
\bibitem [{\citenamefont {Ni}\ \emph {et~al.}(2020)\citenamefont {Ni},
  \citenamefont {Brennecke}, \citenamefont {Gao}, \citenamefont {He},
  \citenamefont {Donsa}, \citenamefont {B\ifmmode~\check{r}\else
  \v{r}\fi{}ezinov\'a}, \citenamefont {He}, \citenamefont {Wu}, \citenamefont
  {Lein}, \citenamefont {Tong},\ and\ \citenamefont
  {Burgd\"orfer}}]{Hongcheng2020}%
  \BibitemOpen
  \bibfield  {author} {\bibinfo {author} {\bibfnamefont {H.~C.}\ \bibnamefont
  {Ni}}, \bibinfo {author} {\bibfnamefont {S.}~\bibnamefont {Brennecke}},
  \bibinfo {author} {\bibfnamefont {X.}~\bibnamefont {Gao}}, \bibinfo {author}
  {\bibfnamefont {P.~L.}\ \bibnamefont {He}}, \bibinfo {author} {\bibfnamefont
  {S.}~\bibnamefont {Donsa}}, \bibinfo {author} {\bibfnamefont
  {I.}~\bibnamefont {B\ifmmode~\check{r}\else \v{r}\fi{}ezinov\'a}}, \bibinfo
  {author} {\bibfnamefont {F.}~\bibnamefont {He}}, \bibinfo {author}
  {\bibfnamefont {J.}~\bibnamefont {Wu}}, \bibinfo {author} {\bibfnamefont
  {M.}~\bibnamefont {Lein}}, \bibinfo {author} {\bibfnamefont {X.~M.}\
  \bibnamefont {Tong}}, \ and\ \bibinfo {author} {\bibfnamefont
  {J.}~\bibnamefont {Burgd\"orfer}},\ }\bibfield  {title} {\enquote {\bibinfo
  {title} {Theory of subcycle linear momentum transfer in strong-field
  tunneling ionization},}\ }\href {\doibase 10.1103/PhysRevLett.125.073202}
  {\bibfield  {journal} {\bibinfo  {journal} {Phys. Rev. Lett.}\ }\textbf
  {\bibinfo {volume} {125}},\ \bibinfo {pages} {073202} (\bibinfo {year}
  {2020})}\BibitemShut {NoStop}%
\bibitem [{\citenamefont {Goulielmakis}(2004)}]{Goulielmakis2004}%
  \BibitemOpen
  \bibfield  {author} {\bibinfo {author} {\bibfnamefont {E}~\bibnamefont
  {Goulielmakis}},\ }\bibfield  {title} {\enquote {\bibinfo {title} {{Direct
  Measurement of Light Waves}},}\ }\href {\doibase 10.1126/science.1100866}
  {\bibfield  {journal} {\bibinfo  {journal} {Science}\ }\textbf {\bibinfo
  {volume} {305}},\ \bibinfo {pages} {1267--1269} (\bibinfo {year}
  {2004})}\BibitemShut {NoStop}%
\bibitem [{\citenamefont {Itatani}\ \emph {et~al.}(2002)\citenamefont
  {Itatani}, \citenamefont {Qu{\'{e}}r{\'{e}}}, \citenamefont {Yudin},
  \citenamefont {Ivanov}, \citenamefont {Krausz},\ and\ \citenamefont
  {Corkum}}]{Itatani2002}%
  \BibitemOpen
  \bibfield  {author} {\bibinfo {author} {\bibfnamefont {J.}~\bibnamefont
  {Itatani}}, \bibinfo {author} {\bibfnamefont {F.}~\bibnamefont
  {Qu{\'{e}}r{\'{e}}}}, \bibinfo {author} {\bibfnamefont {G.~L.}\ \bibnamefont
  {Yudin}}, \bibinfo {author} {\bibfnamefont {M.~Yu}\ \bibnamefont {Ivanov}},
  \bibinfo {author} {\bibfnamefont {F.}~\bibnamefont {Krausz}}, \ and\ \bibinfo
  {author} {\bibfnamefont {P.~B.}\ \bibnamefont {Corkum}},\ }\bibfield  {title}
  {\enquote {\bibinfo {title} {{Attosecond Streak Camera}},}\ }\href {\doibase
  10.1103/PhysRevLett.88.173903} {\bibfield  {journal} {\bibinfo  {journal}
  {Phys. Rev. Lett.}\ }\textbf {\bibinfo {volume} {88}},\ \bibinfo {pages}
  {173903} (\bibinfo {year} {2002})}\BibitemShut {NoStop}%
\bibitem [{\citenamefont {Kienberger}\ \emph {et~al.}(2004)\citenamefont
  {Kienberger}, \citenamefont {Goulielmakis}, \citenamefont {Uiberacker},
  \citenamefont {Baltuska}, \citenamefont {Yakovlev}, \citenamefont {Bammer},
  \citenamefont {Scrinzi}, \citenamefont {Westerwalbesloh}, \citenamefont
  {Kleineberg}, \citenamefont {Heinzmann}, \citenamefont {Drescher},\ and\
  \citenamefont {Krausz}}]{Kienberger2004}%
  \BibitemOpen
  \bibfield  {author} {\bibinfo {author} {\bibfnamefont {R.}~\bibnamefont
  {Kienberger}}, \bibinfo {author} {\bibfnamefont {E.}~\bibnamefont
  {Goulielmakis}}, \bibinfo {author} {\bibfnamefont {M.}~\bibnamefont
  {Uiberacker}}, \bibinfo {author} {\bibfnamefont {A.}~\bibnamefont
  {Baltuska}}, \bibinfo {author} {\bibfnamefont {V.}~\bibnamefont {Yakovlev}},
  \bibinfo {author} {\bibfnamefont {F.}~\bibnamefont {Bammer}}, \bibinfo
  {author} {\bibfnamefont {A.}~\bibnamefont {Scrinzi}}, \bibinfo {author}
  {\bibfnamefont {Th.}\ \bibnamefont {Westerwalbesloh}}, \bibinfo {author}
  {\bibfnamefont {U.}~\bibnamefont {Kleineberg}}, \bibinfo {author}
  {\bibfnamefont {U.}~\bibnamefont {Heinzmann}}, \bibinfo {author}
  {\bibfnamefont {M.}~\bibnamefont {Drescher}}, \ and\ \bibinfo {author}
  {\bibfnamefont {F.}~\bibnamefont {Krausz}},\ }\bibfield  {title} {\enquote
  {\bibinfo {title} {{Atomic transient recorder}},}\ }\href {\doibase
  10.1038/nature02277} {\bibfield  {journal} {\bibinfo  {journal} {Nature}\
  }\textbf {\bibinfo {volume} {427}},\ \bibinfo {pages} {817--821} (\bibinfo
  {year} {2004})}\BibitemShut {NoStop}%
\bibitem [{\citenamefont {Kitzler}\ \emph {et~al.}(2002)\citenamefont
  {Kitzler}, \citenamefont {Milosevic}, \citenamefont {Scrinzi}, \citenamefont
  {Krausz},\ and\ \citenamefont {Brabec}}]{Kitzler2002}%
  \BibitemOpen
  \bibfield  {author} {\bibinfo {author} {\bibfnamefont {M.}~\bibnamefont
  {Kitzler}}, \bibinfo {author} {\bibfnamefont {N.}~\bibnamefont {Milosevic}},
  \bibinfo {author} {\bibfnamefont {A.}~\bibnamefont {Scrinzi}}, \bibinfo
  {author} {\bibfnamefont {F.}~\bibnamefont {Krausz}}, \ and\ \bibinfo {author}
  {\bibfnamefont {T.}~\bibnamefont {Brabec}},\ }\bibfield  {title} {\enquote
  {\bibinfo {title} {{Quantum Theory of Attosecond XUV Pulse Measurement by
  Laser Dressed Photoionization}},}\ }\href {\doibase
  10.1103/PhysRevLett.88.173904} {\bibfield  {journal} {\bibinfo  {journal}
  {Phys. Rev. Lett.}\ }\textbf {\bibinfo {volume} {88}},\ \bibinfo {pages}
  {173904} (\bibinfo {year} {2002})}\BibitemShut {NoStop}%
\bibitem [{\citenamefont {Wang}\ \emph {et~al.}(2010)\citenamefont {Wang},
  \citenamefont {Chini}, \citenamefont {Chen}, \citenamefont {Zhang},
  \citenamefont {He}, \citenamefont {Cheng}, \citenamefont {Wu}, \citenamefont
  {Thumm},\ and\ \citenamefont {Chang}}]{Wang2010}%
  \BibitemOpen
  \bibfield  {author} {\bibinfo {author} {\bibfnamefont {H.}~\bibnamefont
  {Wang}}, \bibinfo {author} {\bibfnamefont {M.}~\bibnamefont {Chini}},
  \bibinfo {author} {\bibfnamefont {S.}~\bibnamefont {Chen}}, \bibinfo {author}
  {\bibfnamefont {C.~H.}\ \bibnamefont {Zhang}}, \bibinfo {author}
  {\bibfnamefont {F.}~\bibnamefont {He}}, \bibinfo {author} {\bibfnamefont
  {Y.}~\bibnamefont {Cheng}}, \bibinfo {author} {\bibfnamefont
  {Y.}~\bibnamefont {Wu}}, \bibinfo {author} {\bibfnamefont {U.}~\bibnamefont
  {Thumm}}, \ and\ \bibinfo {author} {\bibfnamefont {Z.}~\bibnamefont
  {Chang}},\ }\bibfield  {title} {\enquote {\bibinfo {title} {{Attosecond
  Time-Resolved Autoionization of Argon}},}\ }\href {\doibase
  10.1103/PhysRevLett.105.143002} {\bibfield  {journal} {\bibinfo  {journal}
  {Phys. Rev. Lett.}\ }\textbf {\bibinfo {volume} {105}},\ \bibinfo {pages}
  {143002} (\bibinfo {year} {2010})}\BibitemShut {NoStop}%
\bibitem [{\citenamefont {Wickenhauser}\ \emph {et~al.}(2005)\citenamefont
  {Wickenhauser}, \citenamefont {Burgd{\"{o}}rfer}, \citenamefont {Krausz},\
  and\ \citenamefont {Drescher}}]{Wickenhauser2005}%
  \BibitemOpen
  \bibfield  {author} {\bibinfo {author} {\bibfnamefont {Marlene}\ \bibnamefont
  {Wickenhauser}}, \bibinfo {author} {\bibfnamefont {Joachim}\ \bibnamefont
  {Burgd{\"{o}}rfer}}, \bibinfo {author} {\bibfnamefont {Ferenc}\ \bibnamefont
  {Krausz}}, \ and\ \bibinfo {author} {\bibfnamefont {Markus}\ \bibnamefont
  {Drescher}},\ }\bibfield  {title} {\enquote {\bibinfo {title} {{Time Resolved
  Fano Resonances}},}\ }\href {\doibase 10.1103/PhysRevLett.94.023002}
  {\bibfield  {journal} {\bibinfo  {journal} {Phys. Rev. Lett.}\ }\textbf
  {\bibinfo {volume} {94}},\ \bibinfo {pages} {023002} (\bibinfo {year}
  {2005})}\BibitemShut {NoStop}%
\bibitem [{\citenamefont {Ning}\ \emph {et~al.}(2014)\citenamefont {Ning},
  \citenamefont {Peng}, \citenamefont {Song}, \citenamefont {Jiang},
  \citenamefont {Nagele}, \citenamefont {Pazourek}, \citenamefont
  {Burgd\"orfer},\ and\ \citenamefont {Gong}}]{Ning2014}%
  \BibitemOpen
  \bibfield  {author} {\bibinfo {author} {\bibfnamefont {Q.~C.}\ \bibnamefont
  {Ning}}, \bibinfo {author} {\bibfnamefont {L.~Y.}\ \bibnamefont {Peng}},
  \bibinfo {author} {\bibfnamefont {S.~N.}\ \bibnamefont {Song}}, \bibinfo
  {author} {\bibfnamefont {W.~C.}\ \bibnamefont {Jiang}}, \bibinfo {author}
  {\bibfnamefont {S.}~\bibnamefont {Nagele}}, \bibinfo {author} {\bibfnamefont
  {R.}~\bibnamefont {Pazourek}}, \bibinfo {author} {\bibfnamefont
  {J.}~\bibnamefont {Burgd\"orfer}}, \ and\ \bibinfo {author} {\bibfnamefont
  {Q.}~\bibnamefont {Gong}},\ }\bibfield  {title} {\enquote {\bibinfo {title}
  {Attosecond streaking of cohen-fano interferences in the photoionization of
  ${\mathrm{h}}_{2}^{+}$},}\ }\href {\doibase 10.1103/PhysRevA.90.013423}
  {\bibfield  {journal} {\bibinfo  {journal} {Phys. Rev. A}\ }\textbf {\bibinfo
  {volume} {90}},\ \bibinfo {pages} {013423} (\bibinfo {year}
  {2014})}\BibitemShut {NoStop}%
\bibitem [{\citenamefont {Schultze}\ \emph {et~al.}(2010)\citenamefont
  {Schultze}, \citenamefont {Fiess}, \citenamefont {Karpowicz}, \citenamefont
  {Gagnon}, \citenamefont {Korbman}, \citenamefont {Hofstetter}, \citenamefont
  {Neppl}, \citenamefont {Cavalieri}, \citenamefont {Komninos}, \citenamefont
  {Mercouris}, \citenamefont {Nicolaides}, \citenamefont {Pazourek},
  \citenamefont {Nagele}, \citenamefont {Feist}, \citenamefont {Burgdorfer},
  \citenamefont {Azzeer}, \citenamefont {Ernstorfer}, \citenamefont
  {Kienberger}, \citenamefont {Kleineberg}, \citenamefont {Goulielmakis},
  \citenamefont {Krausz},\ and\ \citenamefont {Yakovlev}}]{Schultze2010}%
  \BibitemOpen
  \bibfield  {author} {\bibinfo {author} {\bibfnamefont {M.}~\bibnamefont
  {Schultze}}, \bibinfo {author} {\bibfnamefont {M.}~\bibnamefont {Fiess}},
  \bibinfo {author} {\bibfnamefont {N.}~\bibnamefont {Karpowicz}}, \bibinfo
  {author} {\bibfnamefont {J.}~\bibnamefont {Gagnon}}, \bibinfo {author}
  {\bibfnamefont {M.}~\bibnamefont {Korbman}}, \bibinfo {author} {\bibfnamefont
  {M.}~\bibnamefont {Hofstetter}}, \bibinfo {author} {\bibfnamefont
  {S.}~\bibnamefont {Neppl}}, \bibinfo {author} {\bibfnamefont {A.~L.}\
  \bibnamefont {Cavalieri}}, \bibinfo {author} {\bibfnamefont {Y.}~\bibnamefont
  {Komninos}}, \bibinfo {author} {\bibfnamefont {Th}~\bibnamefont {Mercouris}},
  \bibinfo {author} {\bibfnamefont {C.~A.}\ \bibnamefont {Nicolaides}},
  \bibinfo {author} {\bibfnamefont {R.}~\bibnamefont {Pazourek}}, \bibinfo
  {author} {\bibfnamefont {S.}~\bibnamefont {Nagele}}, \bibinfo {author}
  {\bibfnamefont {J.}~\bibnamefont {Feist}}, \bibinfo {author} {\bibfnamefont
  {J.}~\bibnamefont {Burgdorfer}}, \bibinfo {author} {\bibfnamefont {A.~M.}\
  \bibnamefont {Azzeer}}, \bibinfo {author} {\bibfnamefont {R.}~\bibnamefont
  {Ernstorfer}}, \bibinfo {author} {\bibfnamefont {R.}~\bibnamefont
  {Kienberger}}, \bibinfo {author} {\bibfnamefont {U.}~\bibnamefont
  {Kleineberg}}, \bibinfo {author} {\bibfnamefont {E.}~\bibnamefont
  {Goulielmakis}}, \bibinfo {author} {\bibfnamefont {F.}~\bibnamefont
  {Krausz}}, \ and\ \bibinfo {author} {\bibfnamefont {V.~S.}\ \bibnamefont
  {Yakovlev}},\ }\bibfield  {title} {\enquote {\bibinfo {title} {{Delay in
  Photoemission}},}\ }\href {\doibase 10.1126/science.1189401} {\bibfield
  {journal} {\bibinfo  {journal} {Science}\ }\textbf {\bibinfo {volume}
  {328}},\ \bibinfo {pages} {1658--1662} (\bibinfo {year} {2010})}\BibitemShut
  {NoStop}%
\bibitem [{\citenamefont {Kheifets}\ and\ \citenamefont
  {Ivanov}(2010)}]{Kheifets2010}%
  \BibitemOpen
  \bibfield  {author} {\bibinfo {author} {\bibfnamefont {A.~S.}\ \bibnamefont
  {Kheifets}}\ and\ \bibinfo {author} {\bibfnamefont {I.~A.}\ \bibnamefont
  {Ivanov}},\ }\bibfield  {title} {\enquote {\bibinfo {title} {Delay in atomic
  photoionization},}\ }\href {\doibase 10.1103/PhysRevLett.105.233002}
  {\bibfield  {journal} {\bibinfo  {journal} {Phys. Rev. Lett.}\ }\textbf
  {\bibinfo {volume} {105}},\ \bibinfo {pages} {233002} (\bibinfo {year}
  {2010})}\BibitemShut {NoStop}%
\bibitem [{\citenamefont {Baggesen}\ and\ \citenamefont
  {Madsen}(2010)}]{Baggesen2010}%
  \BibitemOpen
  \bibfield  {author} {\bibinfo {author} {\bibfnamefont {J.~C.}\ \bibnamefont
  {Baggesen}}\ and\ \bibinfo {author} {\bibfnamefont {L.~B.}\ \bibnamefont
  {Madsen}},\ }\bibfield  {title} {\enquote {\bibinfo {title} {Polarization
  effects in attosecond photoelectron spectroscopy},}\ }\href {\doibase
  10.1103/PhysRevLett.104.043602} {\bibfield  {journal} {\bibinfo  {journal}
  {Phys. Rev. Lett.}\ }\textbf {\bibinfo {volume} {104}},\ \bibinfo {pages}
  {043602} (\bibinfo {year} {2010})}\BibitemShut {NoStop}%
\bibitem [{\citenamefont {Ivanov}\ and\ \citenamefont
  {Smirnova}(2011)}]{Ivanov2011}%
  \BibitemOpen
  \bibfield  {author} {\bibinfo {author} {\bibfnamefont {M.}~\bibnamefont
  {Ivanov}}\ and\ \bibinfo {author} {\bibfnamefont {O.}~\bibnamefont
  {Smirnova}},\ }\bibfield  {title} {\enquote {\bibinfo {title} {How accurate
  is the attosecond streak camera?}}\ }\href {\doibase
  10.1103/PhysRevLett.107.213605} {\bibfield  {journal} {\bibinfo  {journal}
  {Phys. Rev. Lett.}\ }\textbf {\bibinfo {volume} {107}},\ \bibinfo {pages}
  {213605} (\bibinfo {year} {2011})}\BibitemShut {NoStop}%
\bibitem [{\citenamefont {Pazourek}\ \emph {et~al.}(2012)\citenamefont
  {Pazourek}, \citenamefont {Feist}, \citenamefont {Nagele},\ and\
  \citenamefont {Burgd\"orfer}}]{Pazourek2012}%
  \BibitemOpen
  \bibfield  {author} {\bibinfo {author} {\bibfnamefont {R.}~\bibnamefont
  {Pazourek}}, \bibinfo {author} {\bibfnamefont {J.}~\bibnamefont {Feist}},
  \bibinfo {author} {\bibfnamefont {S.}~\bibnamefont {Nagele}}, \ and\ \bibinfo
  {author} {\bibfnamefont {J.}~\bibnamefont {Burgd\"orfer}},\ }\bibfield
  {title} {\enquote {\bibinfo {title} {Attosecond streaking of correlated
  two-electron transitions in helium},}\ }\href {\doibase
  10.1103/PhysRevLett.108.163001} {\bibfield  {journal} {\bibinfo  {journal}
  {Phys. Rev. Lett.}\ }\textbf {\bibinfo {volume} {108}},\ \bibinfo {pages}
  {163001} (\bibinfo {year} {2012})}\BibitemShut {NoStop}%
\bibitem [{\citenamefont {Su}\ \emph {et~al.}(2014)\citenamefont {Su},
  \citenamefont {Ni}, \citenamefont {Becker},\ and\ \citenamefont {Jaro\ifmmode
  \acute{n}\else~\'{n}\fi{} Becker}}]{Su2014}%
  \BibitemOpen
  \bibfield  {author} {\bibinfo {author} {\bibfnamefont {J.}~\bibnamefont
  {Su}}, \bibinfo {author} {\bibfnamefont {H.}~\bibnamefont {Ni}}, \bibinfo
  {author} {\bibfnamefont {A.}~\bibnamefont {Becker}}, \ and\ \bibinfo {author}
  {\bibfnamefont {A.}~\bibnamefont {Jaro\ifmmode \acute{n}\else~\'{n}\fi{}
  Becker}},\ }\bibfield  {title} {\enquote {\bibinfo {title}
  {Attosecond-streaking time delays: Finite-range property and comparison of
  classical and quantum approaches},}\ }\href {\doibase
  10.1103/PhysRevA.89.013404} {\bibfield  {journal} {\bibinfo  {journal} {Phys.
  Rev. A}\ }\textbf {\bibinfo {volume} {89}},\ \bibinfo {pages} {013404}
  (\bibinfo {year} {2014})}\BibitemShut {NoStop}%
\bibitem [{\citenamefont {Saalmann}\ and\ \citenamefont
  {Rost}(2020)}]{Saalmann2020}%
  \BibitemOpen
  \bibfield  {author} {\bibinfo {author} {\bibfnamefont {U.}~\bibnamefont
  {Saalmann}}\ and\ \bibinfo {author} {\bibfnamefont {J.~M.}\ \bibnamefont
  {Rost}},\ }\bibfield  {title} {\enquote {\bibinfo {title} {{Proper Time
  Delays Measured by Optical Streaking}},}\ }\href {\doibase
  10.1103/PhysRevLett.125.113202} {\bibfield  {journal} {\bibinfo  {journal}
  {Phys. Rev. Lett.}\ }\textbf {\bibinfo {volume} {125}},\ \bibinfo {pages}
  {113202} (\bibinfo {year} {2020})}\BibitemShut {NoStop}%
\bibitem [{\citenamefont {Pazourek}\ \emph {et~al.}(2013)\citenamefont
  {Pazourek}, \citenamefont {Nagele},\ and\ \citenamefont
  {Burgd{\"{o}}rfer}}]{Pazourek2013}%
  \BibitemOpen
  \bibfield  {author} {\bibinfo {author} {\bibfnamefont {R.}~\bibnamefont
  {Pazourek}}, \bibinfo {author} {\bibfnamefont {S.}~\bibnamefont {Nagele}}, \
  and\ \bibinfo {author} {\bibfnamefont {J.}~\bibnamefont {Burgd{\"{o}}rfer}},\
  }\bibfield  {title} {\enquote {\bibinfo {title} {{Time-resolved photoemission
  on the attosecond scale: Opportunities and challenges}},}\ }\href {\doibase
  10.1039/c3fd00004d} {\bibfield  {journal} {\bibinfo  {journal} {Faraday
  Discussions}\ }\textbf {\bibinfo {volume} {163}},\ \bibinfo {pages}
  {353--376} (\bibinfo {year} {2013})}\BibitemShut {NoStop}%
\bibitem [{\citenamefont {Ivanov}\ \emph {et~al.}(2012)\citenamefont {Ivanov},
  \citenamefont {Kheifets},\ and\ \citenamefont {Serov}}]{Ivanov2012}%
  \BibitemOpen
  \bibfield  {author} {\bibinfo {author} {\bibfnamefont {I.~A.}\ \bibnamefont
  {Ivanov}}, \bibinfo {author} {\bibfnamefont {A.~S.}\ \bibnamefont
  {Kheifets}}, \ and\ \bibinfo {author} {\bibfnamefont {Vladislav~V.}\
  \bibnamefont {Serov}},\ }\bibfield  {title} {\enquote {\bibinfo {title}
  {Attosecond time-delay spectroscopy of the hydrogen molecule},}\ }\href
  {\doibase 10.1103/PhysRevA.86.063422} {\bibfield  {journal} {\bibinfo
  {journal} {Phys. Rev. A}\ }\textbf {\bibinfo {volume} {86}},\ \bibinfo
  {pages} {063422} (\bibinfo {year} {2012})}\BibitemShut {NoStop}%
\bibitem [{\citenamefont {Dixit}\ \emph {et~al.}(2013)\citenamefont {Dixit},
  \citenamefont {Chakraborty},\ and\ \citenamefont {Madjet}}]{Gopal2013}%
  \BibitemOpen
  \bibfield  {author} {\bibinfo {author} {\bibfnamefont {G.}~\bibnamefont
  {Dixit}}, \bibinfo {author} {\bibfnamefont {H.~S.}\ \bibnamefont
  {Chakraborty}}, \ and\ \bibinfo {author} {\bibfnamefont {M.~E.}\ \bibnamefont
  {Madjet}},\ }\bibfield  {title} {\enquote {\bibinfo {title} {Time delay in
  the recoiling valence photoemission of ar endohedrally confined in
  ${\mathrm{c}}_{60}$},}\ }\href {\doibase 10.1103/PhysRevLett.111.203003}
  {\bibfield  {journal} {\bibinfo  {journal} {Phys. Rev. Lett.}\ }\textbf
  {\bibinfo {volume} {111}},\ \bibinfo {pages} {203003} (\bibinfo {year}
  {2013})}\BibitemShut {NoStop}%
\bibitem [{\citenamefont {Serov}\ \emph {et~al.}(2013)\citenamefont {Serov},
  \citenamefont {Derbov},\ and\ \citenamefont {Sergeeva}}]{Serov2013}%
  \BibitemOpen
  \bibfield  {author} {\bibinfo {author} {\bibfnamefont {V.~V.}\ \bibnamefont
  {Serov}}, \bibinfo {author} {\bibfnamefont {V.~L.}\ \bibnamefont {Derbov}}, \
  and\ \bibinfo {author} {\bibfnamefont {T.~A.}\ \bibnamefont {Sergeeva}},\
  }\bibfield  {title} {\enquote {\bibinfo {title} {Interpretation of time delay
  in the ionization of two-center systems},}\ }\href {\doibase
  10.1103/PhysRevA.87.063414} {\bibfield  {journal} {\bibinfo  {journal} {Phys.
  Rev. A}\ }\textbf {\bibinfo {volume} {87}},\ \bibinfo {pages} {063414}
  (\bibinfo {year} {2013})}\BibitemShut {NoStop}%
\bibitem [{\citenamefont {Ossiander}\ \emph {et~al.}(2018)\citenamefont
  {Ossiander}, \citenamefont {Riemensberger}, \citenamefont {Neppl},
  \citenamefont {Mittermair}, \citenamefont {Sch{\"{a}}ffer}, \citenamefont
  {Duensing}, \citenamefont {Wagner}, \citenamefont {Heider}, \citenamefont
  {Wurzer}, \citenamefont {Gerl}, \citenamefont {Schnitzenbaumer},
  \citenamefont {Barth}, \citenamefont {Libisch}, \citenamefont {Lemell},
  \citenamefont {Burgd{\"{o}}rfer}, \citenamefont {Feulner},\ and\
  \citenamefont {Kienberger}}]{Ossiander2018}%
  \BibitemOpen
  \bibfield  {author} {\bibinfo {author} {\bibfnamefont {M.}~\bibnamefont
  {Ossiander}}, \bibinfo {author} {\bibfnamefont {J.}~\bibnamefont
  {Riemensberger}}, \bibinfo {author} {\bibfnamefont {S.}~\bibnamefont
  {Neppl}}, \bibinfo {author} {\bibfnamefont {M.}~\bibnamefont {Mittermair}},
  \bibinfo {author} {\bibfnamefont {M.}~\bibnamefont {Sch{\"{a}}ffer}},
  \bibinfo {author} {\bibfnamefont {A.}~\bibnamefont {Duensing}}, \bibinfo
  {author} {\bibfnamefont {M.~S.}\ \bibnamefont {Wagner}}, \bibinfo {author}
  {\bibfnamefont {R.}~\bibnamefont {Heider}}, \bibinfo {author} {\bibfnamefont
  {M.}~\bibnamefont {Wurzer}}, \bibinfo {author} {\bibfnamefont
  {M.}~\bibnamefont {Gerl}}, \bibinfo {author} {\bibfnamefont {M.}~\bibnamefont
  {Schnitzenbaumer}}, \bibinfo {author} {\bibfnamefont {J.~V.}\ \bibnamefont
  {Barth}}, \bibinfo {author} {\bibfnamefont {F.}~\bibnamefont {Libisch}},
  \bibinfo {author} {\bibfnamefont {C.}~\bibnamefont {Lemell}}, \bibinfo
  {author} {\bibfnamefont {J.}~\bibnamefont {Burgd{\"{o}}rfer}}, \bibinfo
  {author} {\bibfnamefont {P.}~\bibnamefont {Feulner}}, \ and\ \bibinfo
  {author} {\bibfnamefont {R.}~\bibnamefont {Kienberger}},\ }\bibfield  {title}
  {\enquote {\bibinfo {title} {{Absolute timing of the photoelectric
  effect}},}\ }\href {\doibase 10.1038/s41586-018-0503-6} {\bibfield  {journal}
  {\bibinfo  {journal} {Nature}\ }\textbf {\bibinfo {volume} {561}},\ \bibinfo
  {pages} {374--377} (\bibinfo {year} {2018})}\BibitemShut {NoStop}%
\bibitem [{\citenamefont {Zhang}\ and\ \citenamefont
  {Thumm}(2009)}]{Zhang2009}%
  \BibitemOpen
  \bibfield  {author} {\bibinfo {author} {\bibfnamefont {C.-H.}\ \bibnamefont
  {Zhang}}\ and\ \bibinfo {author} {\bibfnamefont {U.}~\bibnamefont {Thumm}},\
  }\bibfield  {title} {\enquote {\bibinfo {title} {Attosecond photoelectron
  spectroscopy of metal surfaces},}\ }\href {\doibase
  10.1103/PhysRevLett.102.123601} {\bibfield  {journal} {\bibinfo  {journal}
  {Phys. Rev. Lett.}\ }\textbf {\bibinfo {volume} {102}},\ \bibinfo {pages}
  {123601} (\bibinfo {year} {2009})}\BibitemShut {NoStop}%
\bibitem [{\citenamefont {Liao}\ and\ \citenamefont {Thumm}(2014)}]{Liao2014}%
  \BibitemOpen
  \bibfield  {author} {\bibinfo {author} {\bibfnamefont {Q.}~\bibnamefont
  {Liao}}\ and\ \bibinfo {author} {\bibfnamefont {U.}~\bibnamefont {Thumm}},\
  }\bibfield  {title} {\enquote {\bibinfo {title} {Attosecond time-resolved
  photoelectron dispersion and photoemission time delays},}\ }\href {\doibase
  10.1103/PhysRevLett.112.023602} {\bibfield  {journal} {\bibinfo  {journal}
  {Phys. Rev. Lett.}\ }\textbf {\bibinfo {volume} {112}},\ \bibinfo {pages}
  {023602} (\bibinfo {year} {2014})}\BibitemShut {NoStop}%
\bibitem [{\citenamefont {Pazourek}\ \emph {et~al.}(2015)\citenamefont
  {Pazourek}, \citenamefont {Nagele},\ and\ \citenamefont
  {Burgd\"orfer}}]{Pazourek2015}%
  \BibitemOpen
  \bibfield  {author} {\bibinfo {author} {\bibfnamefont {R.}~\bibnamefont
  {Pazourek}}, \bibinfo {author} {\bibfnamefont {S.}~\bibnamefont {Nagele}}, \
  and\ \bibinfo {author} {\bibfnamefont {J.}~\bibnamefont {Burgd\"orfer}},\
  }\bibfield  {title} {\enquote {\bibinfo {title} {Attosecond chronoscopy of
  photoemission},}\ }\href {\doibase 10.1103/RevModPhys.87.765} {\bibfield
  {journal} {\bibinfo  {journal} {Rev. Mod. Phys.}\ }\textbf {\bibinfo {volume}
  {87}},\ \bibinfo {pages} {765--802} (\bibinfo {year} {2015})}\BibitemShut
  {NoStop}%
\bibitem [{\citenamefont {{\'{S}}piewanowski}\ and\ \citenamefont
  {Madsen}(2012)}]{Spiewanowski2012}%
  \BibitemOpen
  \bibfield  {author} {\bibinfo {author} {\bibfnamefont {M.~D.}\ \bibnamefont
  {{\'{S}}piewanowski}}\ and\ \bibinfo {author} {\bibfnamefont {L.~B.}\
  \bibnamefont {Madsen}},\ }\bibfield  {title} {\enquote {\bibinfo {title}
  {{Nondipole effects in attosecond photoelectron streaking}},}\ }\href
  {\doibase 10.1103/PhysRevA.86.045401} {\bibfield  {journal} {\bibinfo
  {journal} {Phys. Rev. A}\ }\textbf {\bibinfo {volume} {86}},\ \bibinfo
  {pages} {045401} (\bibinfo {year} {2012})}\BibitemShut {NoStop}%
\bibitem [{\citenamefont {Brennecke}\ and\ \citenamefont
  {Lein}(2018)}]{Brennecke_2018}%
  \BibitemOpen
  \bibfield  {author} {\bibinfo {author} {\bibfnamefont {S.}~\bibnamefont
  {Brennecke}}\ and\ \bibinfo {author} {\bibfnamefont {M.}~\bibnamefont
  {Lein}},\ }\bibfield  {title} {\enquote {\bibinfo {title} {High-order
  above-threshold ionization beyond the electric dipole approximation},}\
  }\href {\doibase 10.1088/1361-6455/aab91f} {\bibfield  {journal} {\bibinfo
  {journal} {J. Phys. B}\ }\textbf {\bibinfo {volume} {51}},\ \bibinfo {pages}
  {094005} (\bibinfo {year} {2018})}\BibitemShut {NoStop}%
\bibitem [{\citenamefont {Liang}\ \emph {et~al.}(2022)\citenamefont {Liang},
  \citenamefont {Zhou}, \citenamefont {Jiang}, \citenamefont {Yu},
  \citenamefont {Li},\ and\ \citenamefont {Lu}}]{Liang2022}%
  \BibitemOpen
  \bibfield  {author} {\bibinfo {author} {\bibfnamefont {J.}~\bibnamefont
  {Liang}}, \bibinfo {author} {\bibfnamefont {Y.}~\bibnamefont {Zhou}},
  \bibinfo {author} {\bibfnamefont {W.~C.}\ \bibnamefont {Jiang}}, \bibinfo
  {author} {\bibfnamefont {M.}~\bibnamefont {Yu}}, \bibinfo {author}
  {\bibfnamefont {M.}~\bibnamefont {Li}}, \ and\ \bibinfo {author}
  {\bibfnamefont {P.}~\bibnamefont {Lu}},\ }\bibfield  {title} {\enquote
  {\bibinfo {title} {Zeeman effect in strong-field ionization},}\ }\href
  {\doibase 10.1103/PhysRevA.105.043112} {\bibfield  {journal} {\bibinfo
  {journal} {Phys. Rev. A}\ }\textbf {\bibinfo {volume} {105}},\ \bibinfo
  {pages} {043112} (\bibinfo {year} {2022})}\BibitemShut {NoStop}%
\bibitem [{\citenamefont {Liang}\ \emph {et~al.}(2021)\citenamefont {Liang},
  \citenamefont {Jiang}, \citenamefont {Liao}, \citenamefont {Ke},
  \citenamefont {Yu}, \citenamefont {Li}, \citenamefont {Zhou},\ and\
  \citenamefont {Lu}}]{Liang:21}%
  \BibitemOpen
  \bibfield  {author} {\bibinfo {author} {\bibfnamefont {J.}~\bibnamefont
  {Liang}}, \bibinfo {author} {\bibfnamefont {W.~C.}\ \bibnamefont {Jiang}},
  \bibinfo {author} {\bibfnamefont {Y.}~\bibnamefont {Liao}}, \bibinfo {author}
  {\bibfnamefont {Q.}~\bibnamefont {Ke}}, \bibinfo {author} {\bibfnamefont
  {M.}~\bibnamefont {Yu}}, \bibinfo {author} {\bibfnamefont {M.}~\bibnamefont
  {Li}}, \bibinfo {author} {\bibfnamefont {Y.}~\bibnamefont {Zhou}}, \ and\
  \bibinfo {author} {\bibfnamefont {P.}~\bibnamefont {Lu}},\ }\bibfield
  {title} {\enquote {\bibinfo {title} {Intensity-dependent angular distribution
  of low-energy electrons generated by intense high-frequency laser pulse},}\
  }\href {\doibase 10.1364/OE.423545} {\bibfield  {journal} {\bibinfo
  {journal} {Opt. Express}\ }\textbf {\bibinfo {volume} {29}},\ \bibinfo
  {pages} {16639} (\bibinfo {year} {2021})}\BibitemShut {NoStop}%
\bibitem [{\citenamefont {Liang}\ \emph {et~al.}(2020)\citenamefont {Liang},
  \citenamefont {Jiang}, \citenamefont {Wang}, \citenamefont {Li},
  \citenamefont {Zhou},\ and\ \citenamefont {Lu}}]{Liang_2020}%
  \BibitemOpen
  \bibfield  {author} {\bibinfo {author} {\bibfnamefont {J.}~\bibnamefont
  {Liang}}, \bibinfo {author} {\bibfnamefont {W.~C.}\ \bibnamefont {Jiang}},
  \bibinfo {author} {\bibfnamefont {S.}~\bibnamefont {Wang}}, \bibinfo {author}
  {\bibfnamefont {M.}~\bibnamefont {Li}}, \bibinfo {author} {\bibfnamefont
  {Y.}~\bibnamefont {Zhou}}, \ and\ \bibinfo {author} {\bibfnamefont
  {P.}~\bibnamefont {Lu}},\ }\bibfield  {title} {\enquote {\bibinfo {title}
  {Atomic dynamic interference in intense linearly and circularly polarized
  {XUV} pulses},}\ }\href {\doibase 10.1088/1361-6455/ab7527} {\bibfield
  {journal} {\bibinfo  {journal} {J. Phys. B}\ }\textbf {\bibinfo {volume}
  {53}},\ \bibinfo {pages} {095601} (\bibinfo {year} {2020})}\BibitemShut
  {NoStop}%
\bibitem [{\citenamefont {Arb\'o}\ \emph {et~al.}(2008)\citenamefont {Arb\'o},
  \citenamefont {Miraglia}, \citenamefont {Gravielle}, \citenamefont
  {Schiessl}, \citenamefont {Persson},\ and\ \citenamefont
  {Burgd\"orfer}}]{Arobo2008}%
  \BibitemOpen
  \bibfield  {author} {\bibinfo {author} {\bibfnamefont {Diego~G.}\
  \bibnamefont {Arb\'o}}, \bibinfo {author} {\bibfnamefont {Jorge~E.}\
  \bibnamefont {Miraglia}}, \bibinfo {author} {\bibfnamefont
  {Mar\'{\i}a~Silvia}\ \bibnamefont {Gravielle}}, \bibinfo {author}
  {\bibfnamefont {Klaus}\ \bibnamefont {Schiessl}}, \bibinfo {author}
  {\bibfnamefont {Emil}\ \bibnamefont {Persson}}, \ and\ \bibinfo {author}
  {\bibfnamefont {Joachim}\ \bibnamefont {Burgd\"orfer}},\ }\bibfield  {title}
  {\enquote {\bibinfo {title} {Coulomb-volkov approximation for near-threshold
  ionization by short laser pulses},}\ }\href {\doibase
  10.1103/PhysRevA.77.013401} {\bibfield  {journal} {\bibinfo  {journal} {Phys.
  Rev. A}\ }\textbf {\bibinfo {volume} {77}},\ \bibinfo {pages} {013401}
  (\bibinfo {year} {2008})}\BibitemShut {NoStop}%
\bibitem [{\citenamefont {Brennecke}\ and\ \citenamefont
  {Lein}(2021)}]{Simon2021}%
  \BibitemOpen
  \bibfield  {author} {\bibinfo {author} {\bibfnamefont {S.}~\bibnamefont
  {Brennecke}}\ and\ \bibinfo {author} {\bibfnamefont {M.}~\bibnamefont
  {Lein}},\ }\bibfield  {title} {\enquote {\bibinfo {title} {Nondipole
  modification of the ac stark effect in above-threshold ionization},}\ }\href
  {\doibase 10.1103/PhysRevA.104.L021104} {\bibfield  {journal} {\bibinfo
  {journal} {Phys. Rev. A}\ }\textbf {\bibinfo {volume} {104}},\ \bibinfo
  {pages} {L021104} (\bibinfo {year} {2021})}\BibitemShut {NoStop}%
\bibitem [{\citenamefont {Jiang}\ and\ \citenamefont {Tian}(2017)}]{Jiang2017}%
  \BibitemOpen
  \bibfield  {author} {\bibinfo {author} {\bibfnamefont {Wei-Chao}\
  \bibnamefont {Jiang}}\ and\ \bibinfo {author} {\bibfnamefont {Xiao-Qing}\
  \bibnamefont {Tian}},\ }\bibfield  {title} {\enquote {\bibinfo {title}
  {{Efficient Split-Lanczos propagator for strong-field ionization of
  atoms}},}\ }\href {\doibase 10.1364/oe.25.026832} {\bibfield  {journal}
  {\bibinfo  {journal} {Optics Express}\ }\textbf {\bibinfo {volume} {25}},\
  \bibinfo {pages} {26832} (\bibinfo {year} {2017})}\BibitemShut {NoStop}%
\bibitem [{\citenamefont {Fr{\"{u}}hling}\ \emph {et~al.}(2009)\citenamefont
  {Fr{\"{u}}hling}, \citenamefont {Wieland}, \citenamefont {Gensch},
  \citenamefont {Gebert}, \citenamefont {Sch{\"{u}}tte}, \citenamefont
  {Krikunova}, \citenamefont {Kalms}, \citenamefont {Budzyn}, \citenamefont
  {Grimm}, \citenamefont {Rossbach}, \citenamefont {Pl{\"{o}}njes},\ and\
  \citenamefont {Drescher}}]{Fruhling2009}%
  \BibitemOpen
  \bibfield  {author} {\bibinfo {author} {\bibfnamefont {U.}~\bibnamefont
  {Fr{\"{u}}hling}}, \bibinfo {author} {\bibfnamefont {M.}~\bibnamefont
  {Wieland}}, \bibinfo {author} {\bibfnamefont {M.}~\bibnamefont {Gensch}},
  \bibinfo {author} {\bibfnamefont {T.}~\bibnamefont {Gebert}}, \bibinfo
  {author} {\bibfnamefont {B.}~\bibnamefont {Sch{\"{u}}tte}}, \bibinfo {author}
  {\bibfnamefont {M.}~\bibnamefont {Krikunova}}, \bibinfo {author}
  {\bibfnamefont {R.}~\bibnamefont {Kalms}}, \bibinfo {author} {\bibfnamefont
  {F.}~\bibnamefont {Budzyn}}, \bibinfo {author} {\bibfnamefont
  {O.}~\bibnamefont {Grimm}}, \bibinfo {author} {\bibfnamefont
  {J.}~\bibnamefont {Rossbach}}, \bibinfo {author} {\bibfnamefont
  {E.}~\bibnamefont {Pl{\"{o}}njes}}, \ and\ \bibinfo {author} {\bibfnamefont
  {M.}~\bibnamefont {Drescher}},\ }\bibfield  {title} {\enquote {\bibinfo
  {title} {{Single-shot terahertz-field-driven X-ray streak camera}},}\ }\href
  {\doibase 10.1038/nphoton.2009.160} {\bibfield  {journal} {\bibinfo
  {journal} {Nat. Photonics}\ }\textbf {\bibinfo {volume} {3}},\ \bibinfo
  {pages} {523--528} (\bibinfo {year} {2009})}\BibitemShut {NoStop}%
\bibitem [{\citenamefont {Schmid}\ \emph {et~al.}(2019)\citenamefont {Schmid},
  \citenamefont {Schnorr}, \citenamefont {Augustin}, \citenamefont {Meister},
  \citenamefont {Lindenblatt}, \citenamefont {Trost}, \citenamefont {Liu},
  \citenamefont {Stojanovic}, \citenamefont {Al-Shemmary}, \citenamefont
  {Golz}, \citenamefont {Treusch}, \citenamefont {Gensch}, \citenamefont
  {K{\"{u}}bel}, \citenamefont {Foucar}, \citenamefont {Rudenko}, \citenamefont
  {Ullrich}, \citenamefont {Schr{\"{o}}ter}, \citenamefont {Pfeifer},\ and\
  \citenamefont {Moshammer}}]{Schmid2019}%
  \BibitemOpen
  \bibfield  {author} {\bibinfo {author} {\bibfnamefont {G.}~\bibnamefont
  {Schmid}}, \bibinfo {author} {\bibfnamefont {K.}~\bibnamefont {Schnorr}},
  \bibinfo {author} {\bibfnamefont {S.}~\bibnamefont {Augustin}}, \bibinfo
  {author} {\bibfnamefont {S.}~\bibnamefont {Meister}}, \bibinfo {author}
  {\bibfnamefont {H.}~\bibnamefont {Lindenblatt}}, \bibinfo {author}
  {\bibfnamefont {F.}~\bibnamefont {Trost}}, \bibinfo {author} {\bibfnamefont
  {Y.}~\bibnamefont {Liu}}, \bibinfo {author} {\bibfnamefont {N.}~\bibnamefont
  {Stojanovic}}, \bibinfo {author} {\bibfnamefont {A.}~\bibnamefont
  {Al-Shemmary}}, \bibinfo {author} {\bibfnamefont {T.}~\bibnamefont {Golz}},
  \bibinfo {author} {\bibfnamefont {R.}~\bibnamefont {Treusch}}, \bibinfo
  {author} {\bibfnamefont {M.}~\bibnamefont {Gensch}}, \bibinfo {author}
  {\bibfnamefont {M.}~\bibnamefont {K{\"{u}}bel}}, \bibinfo {author}
  {\bibfnamefont {L.}~\bibnamefont {Foucar}}, \bibinfo {author} {\bibfnamefont
  {A.}~\bibnamefont {Rudenko}}, \bibinfo {author} {\bibfnamefont
  {J.}~\bibnamefont {Ullrich}}, \bibinfo {author} {\bibfnamefont {C.~D.}\
  \bibnamefont {Schr{\"{o}}ter}}, \bibinfo {author} {\bibfnamefont
  {T.}~\bibnamefont {Pfeifer}}, \ and\ \bibinfo {author} {\bibfnamefont
  {R.}~\bibnamefont {Moshammer}},\ }\bibfield  {title} {\enquote {\bibinfo
  {title} {{Terahertz-Field-Induced Time Shifts in Atomic Photoemission}},}\
  }\href {\doibase 10.1103/PhysRevLett.122.073001} {\bibfield  {journal}
  {\bibinfo  {journal} {Phys. Rev. Lett.}\ }\textbf {\bibinfo {volume} {122}},\
  \bibinfo {pages} {073001} (\bibinfo {year} {2019})}\BibitemShut {NoStop}%
\end{thebibliography}%

%

\end{document}